\documentclass[12pt]{elsarticle}

\usepackage[T1]{fontenc}
\usepackage[utf8]{inputenc}
\usepackage{microtype}
\usepackage{amsmath,amssymb}
\usepackage{siunitx}
\usepackage{natbib} 

\usepackage[colorlinks=true,
            linkcolor=black,  
            citecolor=blue,   
            urlcolor=blue,    
            filecolor=blue]{hyperref}

\usepackage{doi}

\usepackage{tikz}
\usepackage{cleveref}
\usetikzlibrary{arrows.meta,positioning,shapes.geometric,fit,calc}
\usepackage{graphicx}
\usepackage{subcaption} 
\usepackage{float}      

\makeatletter
\def\ps@pprintTitle{%
  \let\@oddhead\@empty
  \let\@evenhead\@empty
  \let\@oddfoot\@empty
  \let\@evenfoot\@oddfoot}
\makeatother

\begin{document}

\begin{frontmatter}

\title{Multi-fidelity Bayesian Optimization Framework for CFD-Based Non-Premixed Burner Design}

\address[ntnu]{Norwegian University of Science and Technology (NTNU), Department of Chemical Engineering, Trondheim, Norway}
\address[icl]{Imperial College London, Process Systems Engineering Group, London, United Kingdom}
\address[senai]{Energy Laboratory, Senai CIMATEC University, Salvador, Brazil}

\author[ntnu]{Patrick Souza Lima}
\author[senai]{Paulo Roberto Santana dos Reis}
\author[senai]{Alex Álisson Bandeira Santos}
\author[icl]{Ehecatl Antonio del Río Chanona\corref{cor2}}
\author[ntnu]{Idelfonso B. R. Nogueira\corref{cor1}}

\cortext[cor1]{Corresponding author. Email: idelfonso.b.d.r.nogueira@ntnu.no (Idelfonso B. R. Nogueira)}
\cortext[cor2]{Corresponding author. Email: a.del-rio-chanona@imperial.ac.uk (A. del Río Chanona)}

\begin{abstract}

We propose a multi-fidelity Bayesian optimization (MF-BO) framework that integrates computational fluid dynamics (CFD) evaluations with Gaussian-process surrogates to efficiently navigate the accuracy--cost trade-off induced by mesh resolution. 
The design vector $\mathbf{x}=[h,l,s]$ (height, length, and mesh element size) defines a continuous fidelity index $Z(h,l,s)$, enabling the optimizer to adaptively combine low- and high-resolution simulations. This framework is applied to a non-premixed burner configuration targeting improved thermal efficiency under hydrogen-enriched fuels. A calibrated runtime model $\hat{t}(h,l,s)$ penalizes computationally expensive queries, while a constrained noisy expected improvement (qNEI) guides sampling under an emissions cap of $2\times10^{-6}$ for \textsc{NO\textsubscript{x}}.

Surrogates trained on CFD  data exhibit stable hyperparameters and physically consistent sensitivities: mean temperature increases with reactor length and fidelity and is weakly negative with height; \textsc{NO\textsubscript{x}} grows with temperature yet tends to decrease with length. The best design achieves $\bar{T}\approx 2.0\times10^{3}\,\mathrm{K}$ while satisfying the \textsc{NO\textsubscript{x}} limit.

Relative to a hypothetical single-fidelity campaign ($Z{=}1$), the MF-BO achieves comparable convergence with $\sim57\%$ lower total wall time by learning the design landscape through fast low-$Z$ evaluations and reserving high-$Z$ CFD for promising candidates. Overall, the methodology offers a generalizable and computationally affordable path for optimizing reacting-flow systems in which mesh-driven fidelity inherently couples accuracy, cost, and emissions. This highlights its potential to accelerate design cycles and reduce resource requirements in industrial burner development and other high-cost CFD-driven applications.
\end{abstract}


\end{frontmatter}

\clearpage 
\section{Introduction}
Energy systems are undergoing a structural transition away from high carbon intensity fuels toward lower carbon carriers and renewable sources. In industry, combustion equipment remains central to reliable heat supply, which motivates pathways that reduce CO$_2$ emissions throughout the life cycle while leveraging existing assets and operational know-how. One pragmatic option is to operate current burners and reactors with hydrogen or hydrogen-enriched natural gas, as well as other low-carbon blends such as bio-syngas or ammonia, thereby reducing carbon intensity at the point of use \citep{IEA_H2_2019,Giacomazzi2023}.

These strategies can accelerate decarbonization, but they simultaneously increase the need for robustness, safety, and efficiency under variable compositions and operating points, with particular attention to NOx emissions. Hydrogen substitution alters flame physics and burner design envelopes: relative to methane, hydrogen exhibits higher laminar flame speed, wider flammability limits, higher diffusivity, and a higher adiabatic flame temperature under comparable conditions. These properties support lean stability and fast heat release, but they also reshape mixing and stabilization mechanisms and increase sensitivity of emissions to injector design, staging, and near-wall flow control \citep{Giacomazzi2023}.

A critical challenge is NOx. Nitrogen oxides are harmful air pollutants linked to respiratory morbidity and environmental impacts such as tropospheric ozone formation, haze, and acid deposition. In high-temperature combustion, NOx often increases with flame temperature through thermal Zeldovich pathways, and hydrogen enrichment tends to raise the adiabatic flame temperature. Unless burner aerodynamics, mixing strategy, and heat-release distribution are adapted, hydrogen use can elevate thermal NOx formation even as CO$_2$ falls \citep{EPA_NO2_2025,Nosek2024}. 

In practical combustor design, a central architectural choice is whether to operate in premixed or non-premixed mode. Fully premixed systems can achieve low emissions at lean conditions, but safe operation with hydrogen-rich fuels is constrained by elevated laminar flame speeds, wider flammability limits, and a strong tendency for flashback into premixers, especially at preheated inlets and high swirl numbers. As a result, many industrial appliances and retrofit pathways favor non-premixed or partially premixed concepts, where fuel and oxidizer are introduced separately to decouple flashback hazards from the fuel train and to retain robust control over staging and heat-release distribution \citep{Aniello2022,Baumgartner2014,Lefebvre2010}. In non-premixed chambers, however, mixture formation occurs inside the reactor and couples multi-scale turbulent transport with finite-rate chemistry and wall-bounded flows, creating strongly inhomogeneous temperature and species fields \citep{Shchepakina2023}. These phenomena lie beyond the fidelity of low-order reactor models such as perfectly stirred or plug-flow representations and therefore require computational fluid dynamics (CFD) with appropriate turbulent-combustion closures (RANS/LES) to capture stabilization mechanisms, local equivalence-ratio fields, and NOx formation pathways \citep{Mastorakos2009,Gicquel2012}.

\begin{figure}[h]
  \centering
  \includegraphics[width=0.9\linewidth]{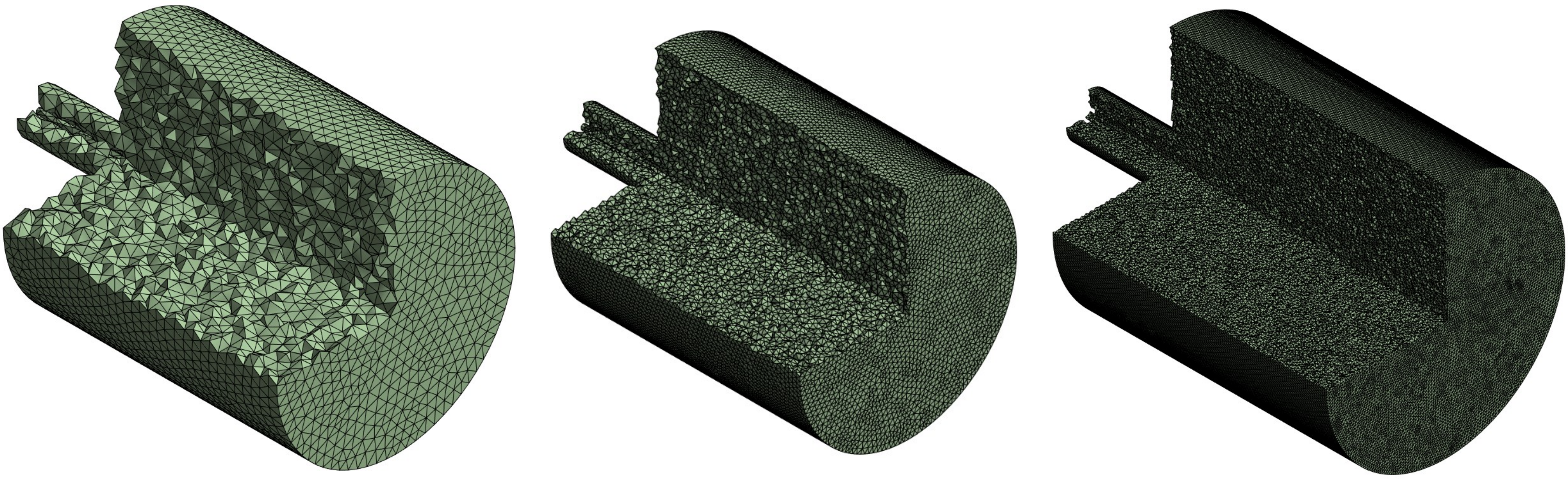}
  \caption{Representative meshes with element sizes of 10~mm, 5~mm, and 2.5~mm. Finer elements increase the number of cells and wall-clock time, highlighting the computational scaling challenge in CFD-based design.}
  \label{fig:mesh}
\end{figure}

In thermo-fluid problems, the practical bottleneck is computational cost. Each CFD evaluation entails a wall-clock time that depends on mesh element size, domain dimensions, solver tolerances, and hardware. As the mesh is refined and numerical criteria are tightened, cost rises rapidly. Consequently, pure manual tuning and dense parametric sweeps are typically too slow and expensive to support systematic design exploration. Figure~\ref{fig:mesh} illustrates how reducing the element size from 10~mm to 5~mm increases the element count and, in turn, the runtime \citep{Vivarelli2025}.

Bayesian optimization has emerged as a sample-efficient paradigm for expensive black-box design, in which a probabilistic surrogate such as a Gaussian process models both response and uncertainty, and an acquisition function guides the next evaluations \citep{Garnett2023}. In contrast to classical design of experiments or local gradient-based search, BO explicitly encodes epistemic uncertainty and uses it to decide where to sample next, which is particularly attractive when each evaluation requires a high-fidelity CFD simulation. Foundational studies have shown that BO can substantially reduce the number of required experiments or simulations across a broad range of engineering applications, from materials and catalysis to control and energy systems \citep{Garnett2023,Shahriari2016,Frazier2018}. In this work, we build on that perspective and view burner geometry and operating conditions as a low-dimensional but highly nonlinear design space where BO can systematically trade off exploration of new regimes against exploitation of promising configurations.

Classical BO formulations, however, often assume either a uniform evaluation cost or a small number of discrete fidelity levels. This assumption underpins many standard acquisition strategies such as Expected Improvement and Upper Confidence Bound, which implicitly treat all new evaluations as equally expensive \citep{Shahriari2016,Frazier2018}. In practice, especially in CFD, this is rarely the case: fidelity is controlled continuously by mesh element size, solver tolerances, and convergence criteria, while geometry and boundary conditions vary with the design itself. As a result, two candidate points with similar physical outputs, for example temperature, concentration or pressure  can differ by an order of magnitude in wall-clock time. Ignoring this heterogeneity can waste computational resources, since some evaluations become disproportionately expensive without commensurate information gain \citep{Savage2023,Lam2015}.

To address this, the multi-fidelity BO literature has proposed a variety of strategies that explicitly account for cost and fidelity. Hierarchical and autoregressive Gaussian process models, as well as cost-weighted acquisition functions, EI per unit cost, have been used to combine information from low- and high-fidelity simulators and to prioritize evaluations that yield the largest expected gain per unit of computational effort \citep{Savage2023,Huang2006}. Recent contributions in process systems engineering have demonstrated how such multi-fidelity and physics-informed BO frameworks can accelerate flowsheet synthesis, reactor optimization, and dynamic operation, often by coupling mechanistic models with data-driven surrogates \citep{LeGratiet2014,Raissi2019}. These studies highlight that BO is not only sample-efficient in a statistical sense, but can also be made resource-aware and aligned with practical computational budgets.

In the context of CFD-based burner design, an additional difficulty arises: fidelity is not a discrete label attached to a fixed simulator, but a continuous quantity that depends on mesh size and domain dimensions, and is therefore intertwined with the design variables themselves. Changing the geometry alters both the physics and the computational cost, because the same nominal element size leads to different cell counts and convergence behaviour. This coupling violates typical assumptions in multi-fidelity BO, where fidelity and design are treated as separable dimensions, and motivates a formulation where fidelity is embedded into the surrogate model and acquisition logic \citep{Savage2024}. 

The present work takes an  step in that direction by introducing a continuous fidelity index derived from mesh resolution and correlating it with both performance (temperature, NOx) and runtime.
In summary, the proposed framework integrates physics-based CFD evaluations, data-driven Gaussian process surrogates, and a cost-aware Bayesian optimization loop to efficiently explore non-premixed burner configurations. By treating mesh-dependent fidelity as a continuous input and explicitly modelling runtime, the method leverages the full flexibility of CFD while mitigating its computational burden, providing a scalable tool for the decarbonization-oriented design of industrial burners.
 
\clearpage
\section{Methodology}

\begin{figure}[H]
\centering
\includegraphics[width=1\linewidth]{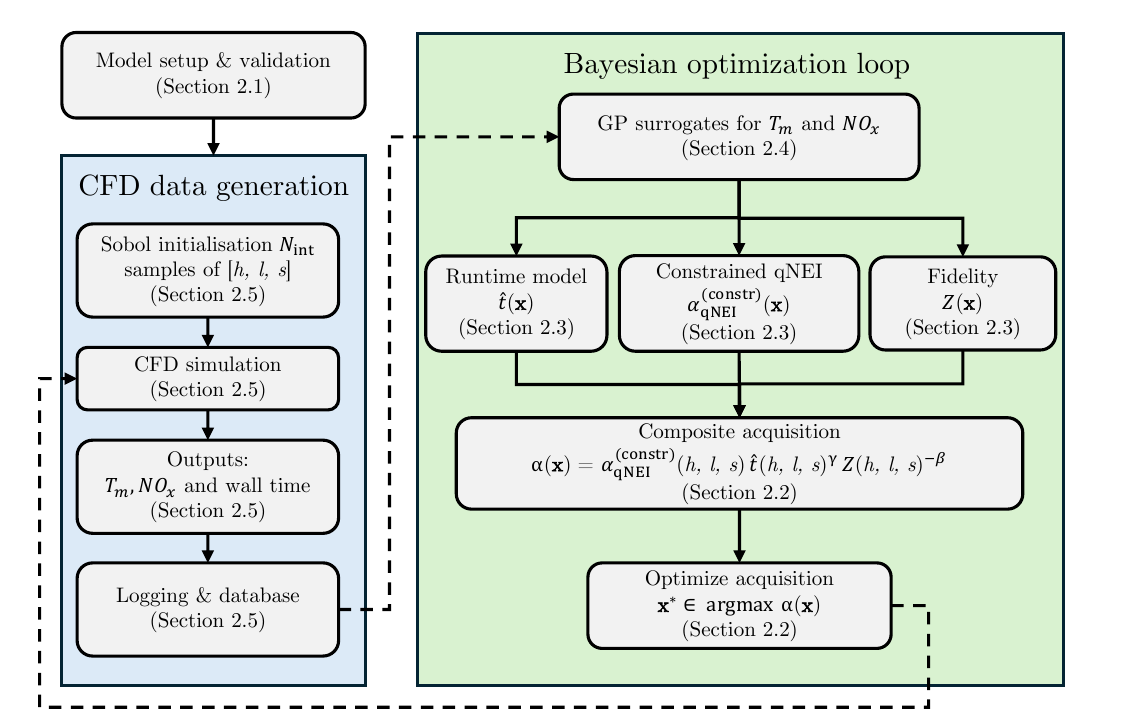}
\caption{The framework couples CFD simulations with Gaussian process surrogates and a cost-aware Bayesian optimization loop. Each candidate geometry $\mathbf{x}=[h,l,s]^{T}$ is simulated in ANSYS, providing $T_m$, $NO_x$, and runtime data used to train Gaussian process models. A calibrated runtime model $\hat{t}(h,l,s)$ and continuous fidelity $Z(h,l,s)$ are integrated into a constrained qNEI acquisition function, which is optimized to select the next candidate.}
\label{fig:methodology_flow}
\end{figure}

Figure~\ref{fig:methodology_flow} provides an overview of the proposed multi-fidelity Bayesian optimization framework.  
The diagram summarizes the coupling between CFD simulations, Gaussian process surrogates, and the cost-aware optimization loop.  
Each block corresponds to one component of the workflow, which will be detailed in the following subsections.  
The figure serves as a roadmap, allowing the reader to visualize the global logic before delving into the specific formulations.

\subsection{Computational fluid dynamic, modeling and validation}
\label{sec:cfd_validation}

A CFD approach was adopted to study non–premixed combustion of binary methane–hydrogen blends under atmospheric conditions. Gravity was included with $g=9.81~\text{m}\,\text{s}^{-2}$. To reduce model complexity and isolate turbulence–chemistry interactions, the configuration was assumed adiabatic with radiation neglected. The flow was modeled as steady and axisymmetric in a pressure-based formulation. Conservation of mass and momentum read
\begin{align}
\nabla\!\cdot(\rho \mathbf{v}) &= 0, \label{eq:continuity}\\
\nabla\!\cdot(\rho \mathbf{v}\mathbf{v}) &= -\nabla p + \nabla\!\cdot\boldsymbol{\tau} + \rho \mathbf{g}, \label{eq:momentum}
\end{align}
where $\rho$ is the density, $\mathbf{v}$ the velocity vector, $p$ the static pressure, and $\boldsymbol{\tau}$ the viscous stress tensor
\begin{equation}
\boldsymbol{\tau} = \mu\!\left[\nabla\mathbf{v} + (\nabla\mathbf{v})^{\!\top}\right] - \frac{2}{3}\mu(\nabla\!\cdot\mathbf{v})\mathbf{I},
\end{equation}
with $\mu$ the molecular viscosity and $\mathbf{I}$ the identity tensor~\citep{Ansys2025}.

Turbulence was treated using the Reynolds average Navier-Stokes(RANS) framework. The Reynolds stress contribution $-\nabla\!\cdot(\rho \overline{\mathbf{v}'\mathbf{v}'})$ was modeled via the standard $k$–$\varepsilon$ closure, which has shown robust performance across reacting flows and engine/turbine-relevant conditions~\citep{Ghadi2024,Yao2025,Ansys2025}. Model constants and wall treatment followed the software defaults unless stated otherwise.

Combustion was described using a non–premixed formulation, with the thermochemical state parameterized by the mixture fraction $f\in[0,1]$, defined from elemental mass fractions as
\begin{equation}
f=\frac{X_i - X_{i,\text{ox}}}{X_{i,\text{fuel}} - X_{i,\text{ox}}},
\end{equation}
where $X_i$ denotes the elemental mass fraction of element $i$, and the subscripts “ox’’ and “fuel’’ refer to oxidizer and fuel inlets, respectively. Detailed chemistry and thermodynamics were represented through the steady diffusion flamelet approach, in which a library of laminar opposed–flow diffusion flame solutions is precomputed over a grid of mixture fraction $f$ and scalar dissipation rate $\chi$. Tabulated quantities such as temperature, enthalpy, and species mass fractions are then retrieved during the CFD run via presumed-PDF averaging~\citep{Ansys2025}. Chemical kinetics followed the GRI–Mech 3.0 mechanism~\citep{Frenklach2024}, which includes 53 species and 325 reactions, and has been extensively validated for CH\textsubscript{4}/H\textsubscript{2} combustion systems~\citep{Wang2025,Xiao2024,Bayramoglu2023}. While the steady flamelet model assumes fast chemistry tending to equilibrium, pollutant formation such as $\mathrm{NO_x}$ occurs over slower timescales. To capture these effects, post-processing source terms for both thermal (Zel’dovich) and prompt (Fenimore) $\mathrm{NO_x}$ were included, consistent with best practices for adiabatic flamelet models~\citep{Ansys2025}.

Pressure–velocity coupling was achieved using the SIMPLE algorithm with Green–Gauss gradient reconstruction. Momentum, pressure, $k$, $\varepsilon$, and species transport equations were discretized with second-order schemes. Convergence criteria of $10^{-4}$ were enforced for flow, turbulence, and PDF equations, and a tighter threshold of $10^{-6}$ for $\mathrm{NO_x}$ transport. All simulations were performed in Ansys Fluent v17.1~\citep{Ansys2025}.

\begin{figure}[H]
    \centering
    \subfloat[\label{fig:mesh_validation_T}Centerline temperature comparison.]{
        \includegraphics[width=0.47\linewidth]{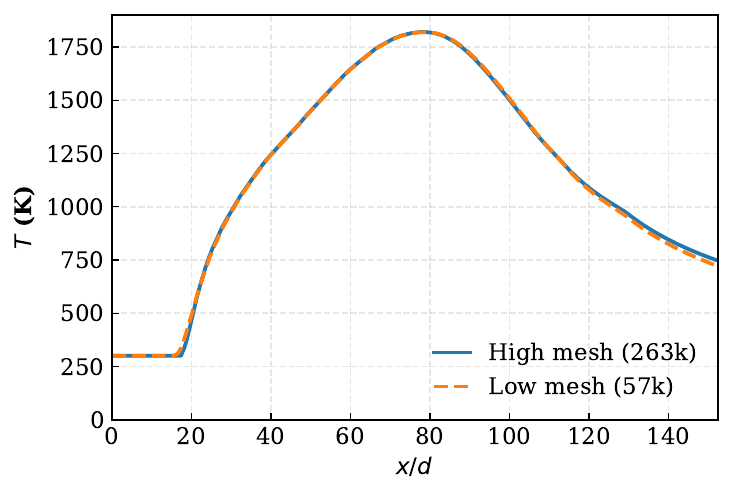}}
    \hfill
    \subfloat[\label{fig:mesh_validation_NOx}Centerline $\mathrm{NO_x}$ mass fraction comparison.]{
        \includegraphics[width=0.47\linewidth]{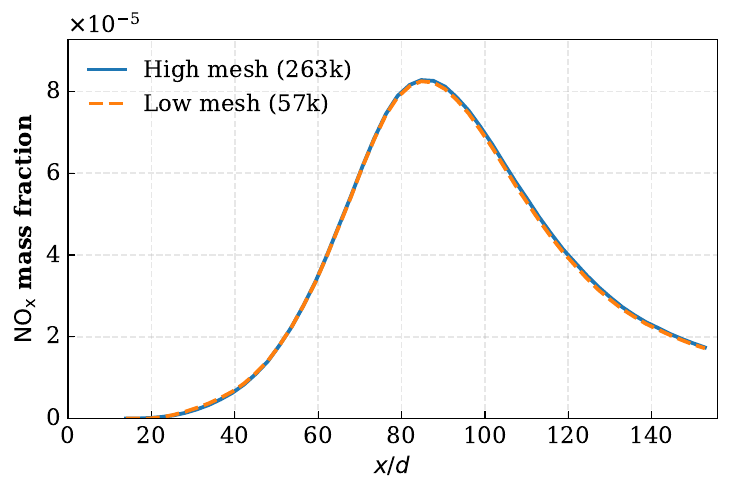}}
    \caption{Mesh-independence validation for the 2D axisymmetric Sandia D simulation.
    (a) The temperature field shows negligible differences between the coarse (57k) and fine (263k) grids, confirming mesh convergence for scalar and thermal fields.
    (b) The $\mathrm{NO_x}$ mass fraction distribution exhibits similarly consistent profiles,
    with deviations below 2\% across the flame region.}
    \label{fig:mesh_validation}
\end{figure}

A mesh-independence study was performed using a 2D axisymmetric domain to verify
spatial discretization effects. Two grids, with 57,175 and 263,001 elements,
were compared for methane/air combustion. Axial profiles of temperature and
$\mathrm{NO_x}$ mass fraction showed close agreement between both meshes, confirming
that the coarser grid provided sufficient resolution. The final mesh featured
targeted refinement along the jet core, inlet regions, and initial reaction
zone to capture steep scalar gradients, as illustrated in Fig.~\ref{fig:mesh_validation}.

\begin{figure}[H]
    \centering
    \subfloat[\label{fig:SandiaD_T}Centerline temperature.]{
        \includegraphics[width=0.48\linewidth]{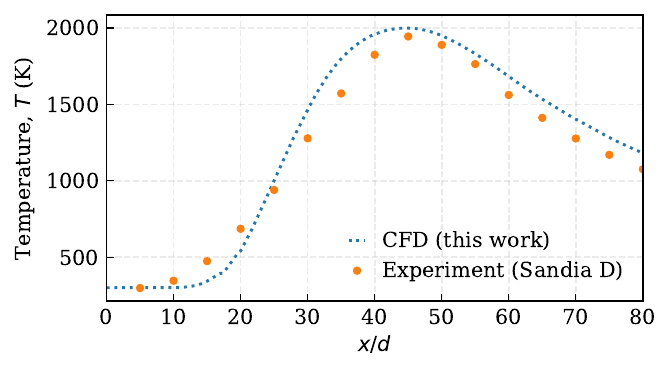}}
    \hfill
    \subfloat[\label{fig:SandiaD_NOx}Centerline $\mathrm{NO_x}$ mass fraction.]{
        \includegraphics[width=0.48\linewidth]{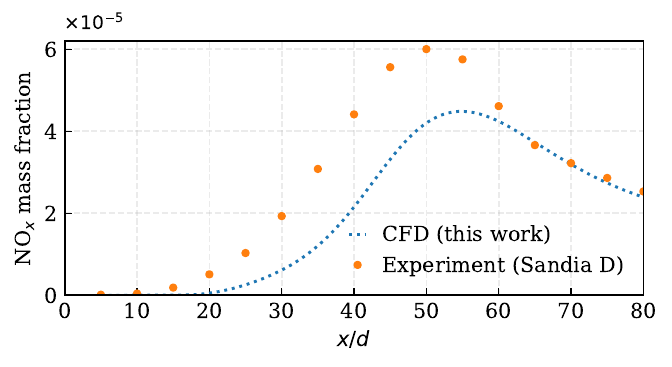}}
    \caption{Validation of the CFD model against Sandia Flame D. 
    (a) Temperature: average deviation of 10.6\%; mild overprediction at the burner exit and underprediction for $x/d<25$. 
    (b) $\mathrm{NO_x}$: underestimated up to $x/d\approx63$, with exit-plane deviation below 6\%. 
    The results support the quantitative reliability of the 2D axisymmetric steady flamelet model with $k$--$\varepsilon$ turbulence and GRI--Mech~3.0.}
    \label{fig:SandiaD_validation}
\end{figure}

The validated numerical model was applied to the canonical Sandia Flame D configuration~\citep{Barlow2005,Bayramoglu2023}. Figure~\ref{fig:SandiaD_validation} compares centerline temperature and $\mathrm{NO_x}$ mass fraction with experiments, showing overall good agreement along the centerline ($d=7.2$~mm is the fuel-nozzle diameter).

 The temperatures are slightly overpredicted near the exit and mildly underpredicted for $x/d<25$, consistent with a lift-off position shifted downstream. Predicted $\mathrm{NO_x}$ is lower than measured up to $x/d\approx63$ but aligns well downstream, yielding an exit-plane error below 6\%. These discrepancies are plausibly linked to the adiabatic and no-radiation assumptions adopted here, yet the overall agreement confirms the model’s suitability for the operating range of interest~\citep{Ansys2025,Frenklach2024,Wang2025,Xiao2024}.

\begin{figure}[H]
    \centering
    \begin{subfigure}[t]{0.48\textwidth}
        \centering
        \includegraphics[width=\textwidth]{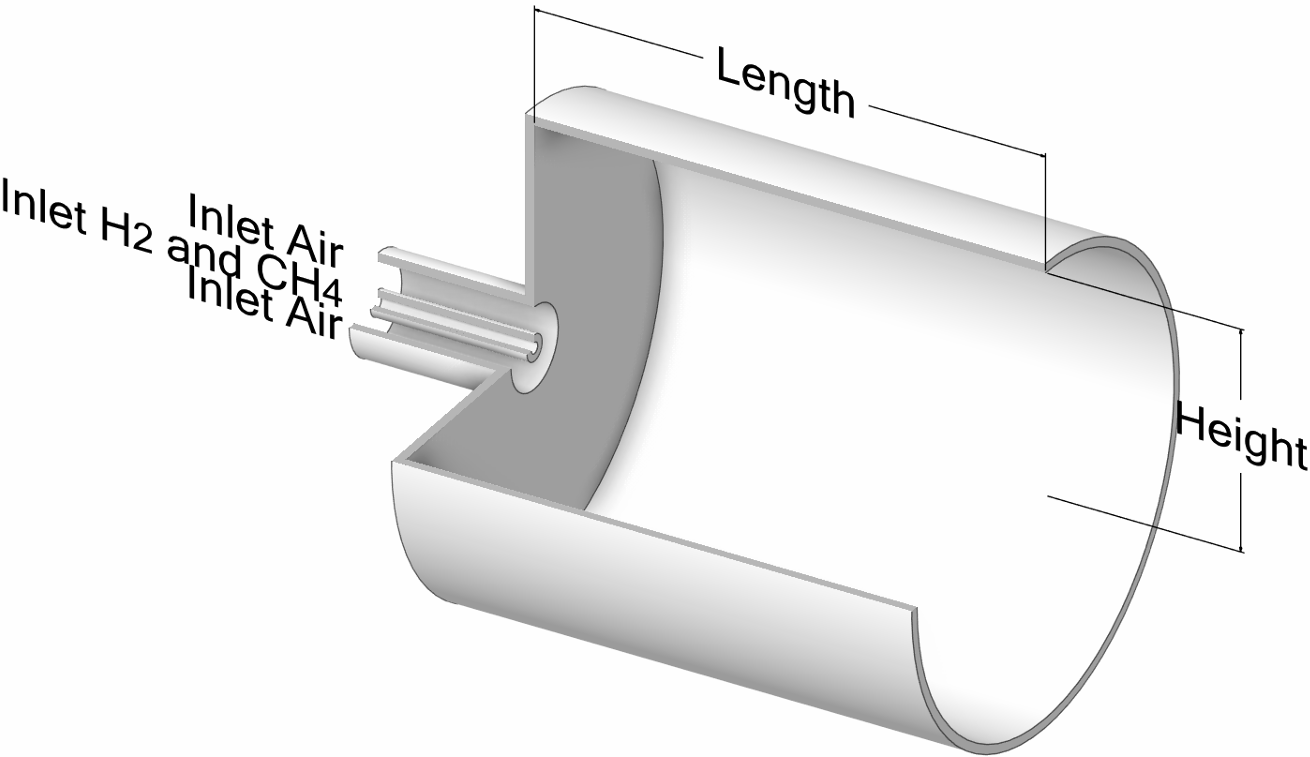}
        \caption{Three-dimensional burner geometry without mixture domain highlighted.}
        \label{fig:geom_nomix}
    \end{subfigure}
    \hfill
    \begin{subfigure}[t]{0.48\textwidth}
        \centering
        \includegraphics[width=\textwidth]{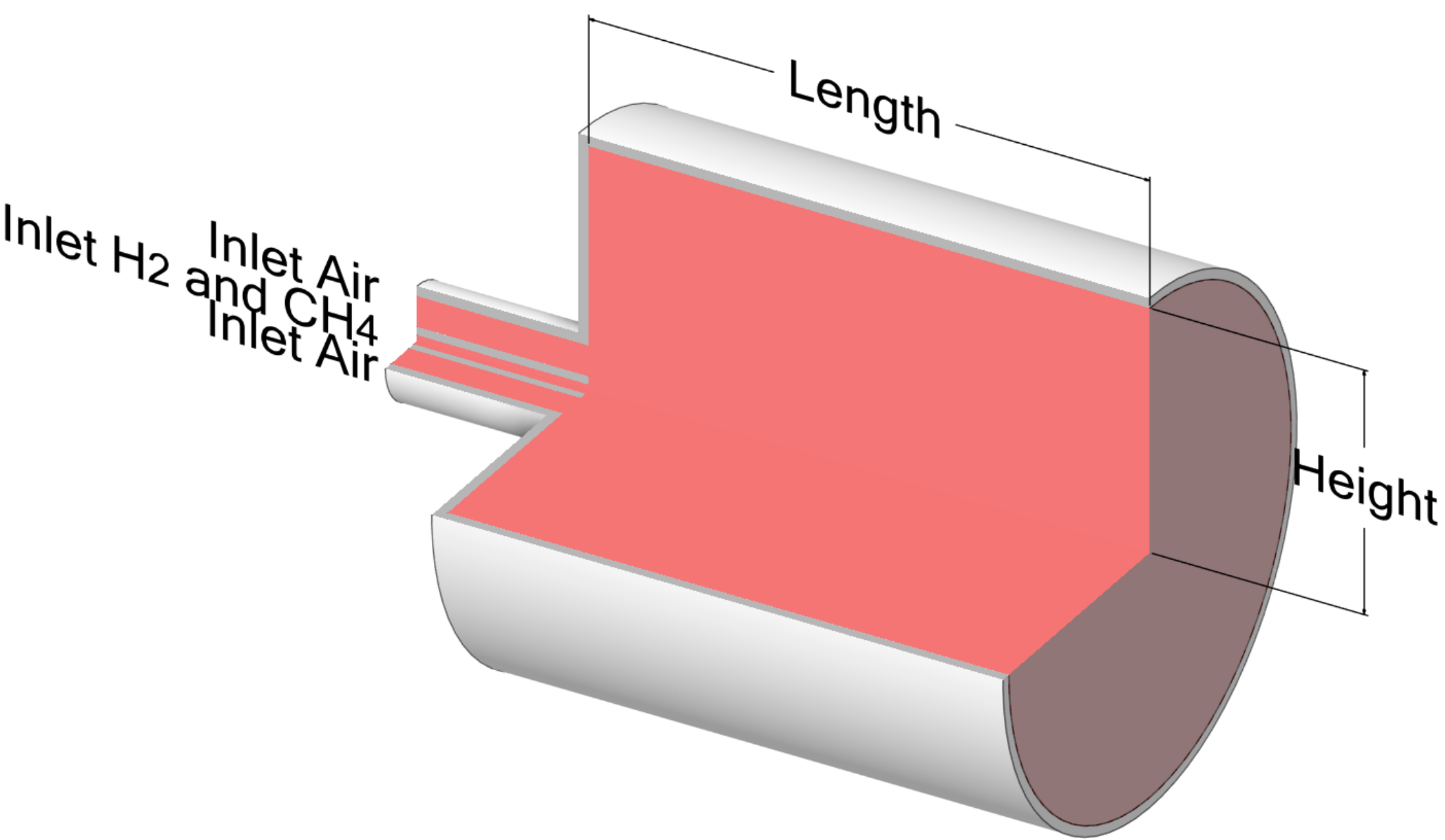}
        \caption{Three-dimensional burner geometry showing the fluid mixture domain (in red).}
        \label{fig:geom_mix}
    \end{subfigure}
    
    \vspace{4mm}
    \begin{subfigure}[t]{0.7\textwidth}
        \centering
        \includegraphics[width=\textwidth]{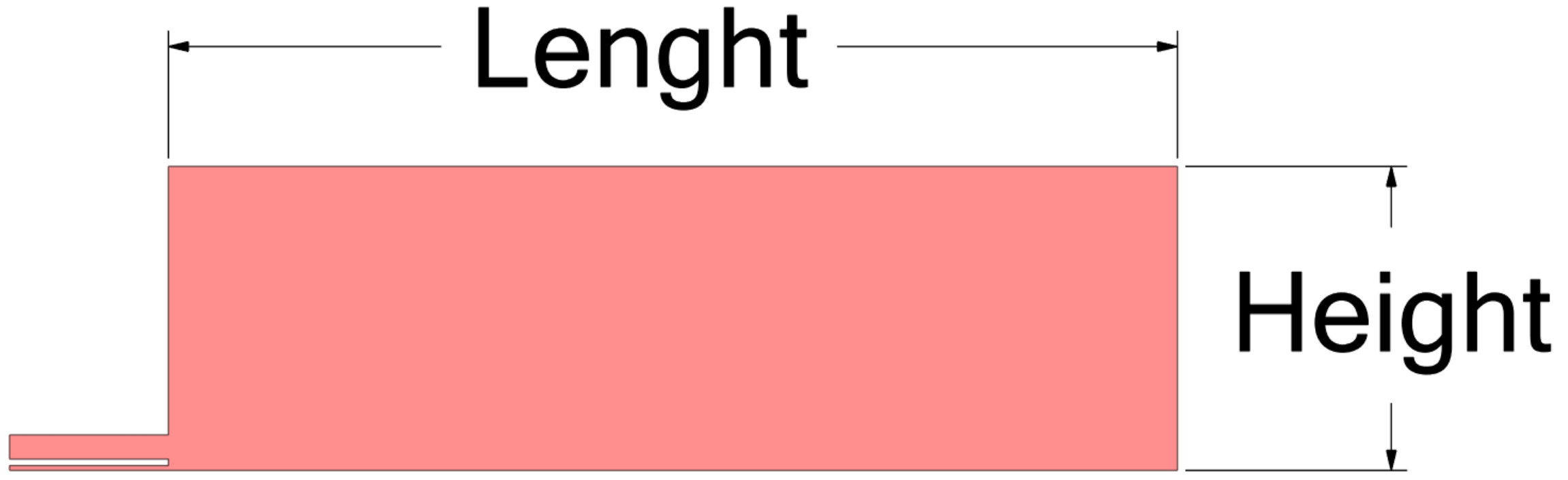}
        \caption{Two-dimensional axisymmetric simplification of the computational domain ($h \times l$).}
        \label{fig:geom_plane}
    \end{subfigure}

    \caption{Computational domain of the confined non–premixed burner used for CFD simulations. 
    Panels (a) and (b) depict the three-dimensional burner geometry prior to simplification: 
    (a) shows the burner structure with its coaxial inlets for air and premixed \(\mathrm{H_2/CH_4}\) fuel, 
    while (b) highlights the internal region filled by the fluid mixture domain (in red). 
    Panel (c) presents the corresponding two-dimensional axisymmetric configuration adopted 
    during the validation and optimization stages. The geometric design variables, height (\(h\)) 
    and length (\(l\)), were parameterized in this simplified 2D plane.}
    \label{fig:geometry_combined}
\end{figure}

Given its validated predictive capability and favorable cost–accuracy balance, the same two-dimensional axisymmetric modeling strategy (Figure~\ref{fig:geometry_combined}\subref*{fig:geom_plane}) was adopted for the subsequent optimization campaign. This approach ensured physical consistency across the explored design space while allowing a substantial reduction in computational cost per design compared to a full three-dimensional formulation (Figures~\ref{fig:geometry_combined}\subref*{fig:geom_nomix}–\subref*{fig:geom_mix}).

The computational domain, illustrated in Figure~\ref{fig:geometry_combined}, represents a confined non–premixed burner with concentric air and fuel inlets. The geometry comprises two coaxial entries: an outer air inlet of width \(37.455~\text{mm}\) and an inner fuel inlet of \(6~\text{mm}\), both aligned along the burner axis. Separate injection of the two streams ensures operational safety and reproduces typical industrial configurations. The air velocity was set to \(0.712~\text{m/s}\), while the fuel jet entered at \(12~\text{m/s}\), corresponding to a premixed composition of \(20\%\ \text{H}_2\) and \(80\%\ \text{CH}_4\) by volume.

At the outlet, a pressure boundary condition of \(p_\text{out}=1~\text{atm}\) was imposed, with zero diffusive flux for all transported variables. No-slip and adiabatic conditions were applied along the burner walls. The selected domain height and length ensured full development of the flame and flow fields while minimizing unnecessary computational volume. These boundary conditions were consistently applied in both the validation and optimization stages to guarantee comparable flow and reaction regimes.

\subsection{Problem formulation and design variables}

In the proposed Bayesian optimization framework, each new candidate design is selected by maximizing a cost- and constraint-aware acquisition function that balances expected performance, emission feasibility, and computational cost. At iteration $k$, the next design $\mathbf{x}^*$ is obtained as
\begin{equation}
\mathbf{x}^*\in \operatorname*{arg\,max}_{\mathbf{x} = [h,\,l,\,s]^{\mathrm{T}}} 
\; \alpha_{\text{qNEI}}^{(\text{constr})}(\mathbf{x}) \;
z(\mathbf{x})^{\gamma} \;
\hat{t}(\mathbf{x})^{-\beta},
\label{eq:acquisition_penalized}
\end{equation}

where $\alpha_{\text{qNEI}}^{(\text{constr})}(\mathbf{x})$ is the constrained noisy expected improvement, $z(h,l,s)$ is a normalized fidelity indicator, and $\hat{t}(h,l,s)$ is an analytical runtime model. The exponents $\beta$ and $\gamma$ control the trade-off between computational cost, fidelity, and expected information gain. This formulation makes explicit that the optimizer seeks designs that are promising in terms of mean reactor temperature, likely to satisfy the NO$_x$ constraint, and computationally affordable.

The design vector is defined as
\[
\mathbf{x} = [h,\,l,\,s]^{\mathrm{T}},
\]
where three design variables are considered: the burner height $h$, the burner length $l$, and the characteristic mesh element size $s$, which acts as a continuous fidelity parameter in the CFD simulations. The variables $h$ and $l$ define the geometric domain and directly influence the flame structure, mixing efficiency, and residence time, whereas $s$ controls the spatial resolution of the numerical grid and, consequently, the computational cost and level of physical detail captured in each simulation.

Each candidate design $\mathbf{x}$ is evaluated through a CFD simulation that returns the mean reactor temperature $T_m(\mathbf{x})$ and the corresponding concentration $NO_x(\mathbf{x})$. From a physical standpoint, the underlying engineering problem can be stated as maximizing $T_m$ while enforcing an upper bound on NO$_x$ emissions. The corresponding constrained optimization problem is
\begin{equation}
\begin{aligned}
\max_{\mathbf{x}=[h,l,s]^{\mathrm{T}}} \quad & T_m(\mathbf{x}) \\
\text{s.t.} \quad & NO_x(\mathbf{x}) \le NO_{x,\max}, \\
& h_{\min} \le h \le h_{\max}, \\
& l_{\min} \le l \le l_{\max}, \\
& s_{\min} \le s \le s_{\max}.
\end{aligned}
\label{eq:opt_problem}
\end{equation}
where $NO_{x,\max} = 2\times10^{-6}$ denotes the emission threshold. The search space is defined within physically meaningful bounds obtained from the CFD domain configuration and design constraints: $h \in [100,\,250]~\text{mm}$, $l \in [250,\,1500]~\text{mm}$, and $s \in [0.35,\,3.0]~\text{mm}$.

These ranges were selected to remain consistent with the geometric configurations and mesh resolutions adopted during the experimental validation of the CFD model, ensuring that the optimization process explores only regions of the design space that were previously verified to produce reliable numerical predictions. The element size range, in particular, was determined based on the mesh sensitivity analysis performed as part of the validation campaign, balancing computational cost and numerical accuracy.

\subsection{Constrained acquisition, fidelity parametrization and runtime model}

Equation~\eqref{eq:acquisition_penalized} shows that the Bayesian optimizer combines three main ingredients: (i) the expected improvement in mean temperature with the probability of satisfying the NO$_x$ constraint, (ii) penalties or incentives related to mesh fidelity and (iii) other penalties or incentives to runtime. This subsection details how each of these terms is constructed.

The feasibility of each candidate design with respect to the NO$_x$ constraint is evaluated probabilistically using the posterior distribution of the Gaussian process surrogate for NO$_x$. Assuming a normal predictive distribution 
\[
NO_x(\mathbf{x}) \sim \mathcal{N}\!\big(\mu_{NO_x}(\mathbf{x}), \sigma_{NO_x}^2(\mathbf{x})\big),
\]
the probability of feasibility is computed as
\begin{equation}
P_{\text{feasible}}\!\left(NO_x(\mathbf{x}) \le NO_{x,\max}\right)
= 
\Phi\!\left(
\frac{NO_{x,\max} - \mu_{NO_x}(\mathbf{x})}{\sigma_{NO_x}(\mathbf{x})}
\right),
\label{eq:prob_feasible}
\end{equation}
where $\Phi(\cdot)$ is the cumulative distribution function of the standard normal distribution. This probability is used as a multiplicative weight in the constrained noisy expected improvement, such that the expected gain is weighted by the likelihood that a candidate satisfies the emission limit. The constrained acquisition is thus given by
\begin{equation}
\alpha_{\text{qNEI}}^{(\text{constr})}(\mathbf{x})
=
\mathbb{E}\!\left[
\max\!\left(T_m(\mathbf{x}) - T_m^{\text{best}},\, 0\right)
\right]
P_{\text{feasible}}\!\left(NO_x(\mathbf{x}) \le NO_{x,\max}\right),
\label{eq:constrained_qnei}
\end{equation}
where $T_m^{\text{best}}$ is the best temperature observed so far. The expected improvement term quantifies the potential gain in the objective, while the probability factor enforces soft feasibility in a probabilistic sense.

In CFD-based optimization, the computational cost of each simulation depends strongly on the mesh resolution, which determines both the number of control volumes and the convergence rate of the solver. In this work, the mesh element size $s$ is treated as a continuous fidelity parameter that directly influences the level of numerical detail and the associated wall-clock time. To describe this dependency, an adaptive analytical runtime model $\hat{t}(h,l,s)$ is employed. Its parameters are initially estimated from the wall-clock times collected during the design of experiments (DOE) stage and subsequently updated every ten optimization iterations to reflect the runtime behavior observed throughout the search process. This dynamic calibration enables the model to capture nonlinear cost trends that may arise from varying geometries and solver conditions.

The analytical expression adopted for the runtime model is
\begin{equation}
\hat{t}(h,l,s) = C + A_1 \exp(B_1\,z(h,l,s)) + A_2 \exp\!\left[B_2\,\frac{(h\,l)}{A_0}\right],
\label{eq:time_model}
\end{equation}
where $C$, $A_1$, $A_2$, $B_1$, $B_2$, and $A_0$ are coefficients continuously refined from measured solver runtimes. The first exponential term captures the sensitivity of runtime to mesh density, represented by the normalized fidelity variable $z$, while the second term accounts for the geometric influence of the reactor area $h\,l$ on the total cell count and computational load.

The fidelity variable $z(h,l,s)$ is defined from the local mesh density $\rho = (1/s)/(h\,l)$, normalized between the lower and upper bounds corresponding to the coarsest and finest meshes tested, $s_{\max}$ and $s_{\min}$, respectively:
\begin{equation}
z(h,l,s) = \frac{\ln\!\left(\rho/\rho_{\min}\right)}{\ln\!\left(\rho_{\max}/\rho_{\min}\right)}, 
\quad 
\rho_{\min} = \frac{1/s_{\max}}{h\,l}, \quad 
\rho_{\max} = \frac{1/s_{\min}}{h\,l}.
\label{eq:z_fidelity}
\end{equation}
This normalization maps the mesh density onto the range $z \in [0,1]$, where $z=0$ represents the coarsest admissible mesh and $z=1$ corresponds to the highest fidelity. This formulation allows the optimizer to continuously trade off solution accuracy against computational expense without requiring discrete fidelity levels.

\begin{figure}[H]
    \centering
    \includegraphics[width=0.85\linewidth]{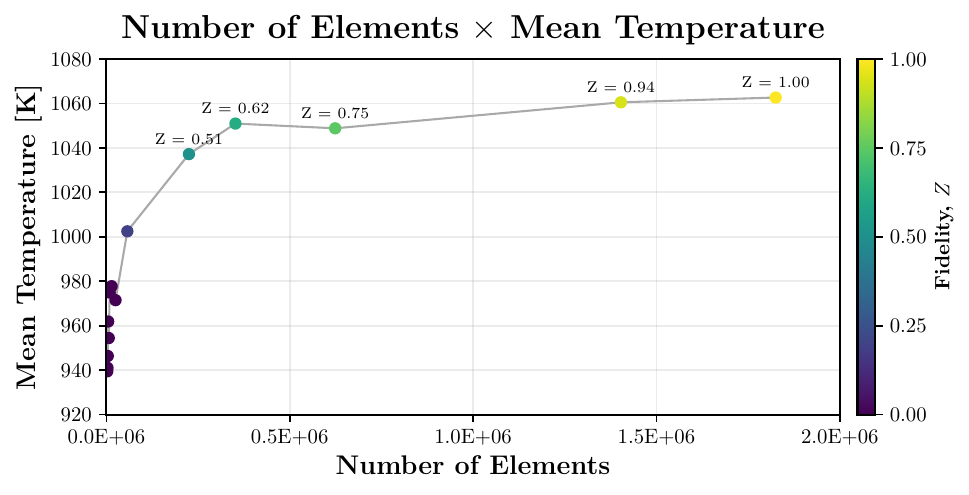}
    \caption{Mesh sensitivity analysis used to calibrate the fidelity bounds. 
Each point corresponds to a CFD simulation with a given element size $s$, 
showing the mean reactor temperature as a function of the total number of elements. 
The color scale represents the normalized fidelity $Z$, obtained from Eq.~\eqref{eq:z_fidelity}. 
The limits $s_{\max}=3.0~\text{mm}$ and $s_{\min}=0.35~\text{mm}$ were selected to encompass 
the coarse and fine mesh extremes that produced numerically convergent and physically consistent 
results for a representative domain of $l = 1000~\text{mm}$ and $h = 200~\text{mm}$. 
These reference cases define the lower and upper bounds for fidelity normalization 
during Bayesian optimization.}
    \label{fig:mesh_fidelity}
\end{figure}

The upper and lower bounds of the element size, $s_{\min}=0.35~\text{mm}$ and $s_{\max}=3.0~\text{mm}$, 
were determined from a dedicated mesh refinement study. 
As illustrated in Figure~\ref{fig:mesh_fidelity}, mesh refinement beyond $s_{\min}$ produced 
negligible variation in the mean temperature ($<1\%$), indicating grid independence, 
while coarsening beyond $s_{\max}$ led to visible numerical diffusion and unstable convergence. 
These limits therefore represent a physically and computationally consistent range for fidelity 
parameterization within the optimization workflow.

Within the Bayesian optimization framework, combining Eqs.~\eqref{eq:constrained_qnei}, \eqref{eq:time_model}, and \eqref{eq:z_fidelity}  yields a fully cost- and constraint-aware acquisition criterion. This formulation directs the optimizer to prioritize evaluations that are both feasible and computationally efficient, adaptively selecting geometry and mesh resolution throughout the search. The next subsection details the construction of the Gaussian process surrogates used to evaluate $\alpha_{\text{qNEI}}^{(\text{constr})}(\mathbf{x})$ and to quantify the predictive uncertainty of $T_m$ and $NO_x$.

\subsection{Surrogate modeling with Gaussian processes}

The high computational cost of CFD evaluations motivates the use of probabilistic surrogate models to approximate the system response with quantified uncertainty. In this work, Gaussian processes (GPs) are adopted as surrogate models for the mean reactor temperature \(T_m\) and for the  \(NO_x\), providing smooth, differentiable approximations that enable Bayesian inference and sample-efficient optimization. The GP prior defines a distribution over possible response functions, updated through Bayesian conditioning as new CFD evaluations become available.

Each response variable is modeled by an independent GP, implemented in BoTorch and GPyTorch \citep{botorch2020,gpyTorch2018}. The temperature model \(m_{T_m}\) is trained to predict the scalar response \(T_m(\mathbf{x})\), which serves as the optimization objective, while the emission model \(m_{NO_x}\) predicts \(NO_x(\mathbf{x})\) and defines a probabilistic feasibility constraint. The two surrogates are subsequently combined into a joint multi-output GP structure to enable correlated acquisition and consistent uncertainty propagation.

Both models employ a Matérn kernel with smoothness parameter \(\nu = 2.5\) and automatic relevance determination (ARD) across the three input dimensions \([h,\,l,\,s]\). The covariance structure of the kernel is expressed as

\begin{equation}
k(\mathbf{x}_i, \mathbf{x}_j)
= \sigma_f^2
\left(1 + \frac{\sqrt{5}\,r_{ij}}{\ell} + \frac{5\,r_{ij}^2}{3\ell^2}\right)
\exp\!\left(-\frac{\sqrt{5}\,r_{ij}}{\ell}\right),
\label{eq:matern25}
\end{equation}

where \(r_{ij} = \|\mathbf{x}_i - \mathbf{x}_j\|_2\) is the Euclidean distance between inputs, \(\ell = [\ell_h, \ell_l, \ell_s]\) are the lengthscales associated with each variable, and \(\sigma_f^2\) is the output-scale variance. This formulation balances local flexibility with global smoothness, allowing the surrogate to capture both broad geometric trends and localized nonlinear effects. 

Gamma priors are imposed on the kernel hyperparameters to regularize the training and prevent overfitting: the lengthscales \(\ell_d\) follow \(\text{Gamma}(3.0, 6.0)\), and the output-scale prior follows \(\text{Gamma}(2.0, 0.15)\). The Gaussian likelihood variance \(\sigma_n^2\) is constrained by a lower bound of \(10^{-4}\) through a positivity constraint to ensure numerical stability. Input variables are normalized to the unit hypercube \([0,1]^3\) via an affine transform, and outputs are standardized to zero mean and unit variance to improve conditioning during hyperparameter optimization.

The model parameters are trained by maximizing the exact marginal log-likelihood (MLL) under the assumption of Gaussian noise. The MLL objective is given by

\begin{equation}
\mathcal{L}_{\text{MLL}} 
= 
- \frac{1}{2}\mathbf{y}^\top K^{-1}\mathbf{y}
- \frac{1}{2}\log |K|
- \frac{n}{2}\log(2\pi),
\label{eq:mll}
\end{equation}

where \(\mathbf{y}\) is the vector of training outputs, and \(K = K(X,X) + \sigma_n^2 I\) is the covariance matrix evaluated at the input set \(X\). The hyperparameters \(\{\ell, \sigma_f, \sigma_n\}\) are optimized using gradient-based algorithms with automatic differentiation. Warm-starting is applied at each iteration by reusing the hyperparameter state from the previous GP fit, which accelerates convergence and improves model continuity throughout the optimization process.

The resulting models provide not only mean predictions but also posterior variances that quantify epistemic uncertainty. These uncertainties are exploited by the acquisition function to balance exploration and exploitation, allowing the Bayesian optimization loop to target regions with both high expected performance and limited prior information. This probabilistic representation of model confidence is key to achieving sample-efficient design optimization under high computational cost.

\subsection{Implementation and computational workflow}

The proposed Bayesian optimization framework was implemented entirely in Python using the BoTorch and GPyTorch libraries \citep{botorch2020,gpyTorch2018}, which provide modular interfaces for probabilistic surrogate modeling and gradient-based acquisition optimization. The optimization loop communicates directly with ANSYS Workbench through journal automation scripts (\texttt{.wbjn}) executed via the \texttt{RunWB2.exe} interface. Each candidate configuration, defined by the variables \([h,\,l,\,s]\), is written to an input file (\texttt{in.json}) and passed to the CFD solver, which returns the mean reactor temperature \(T_m\), the NO$_x$ concentration, and auxiliary quantities such as wall-clock time, mesh size, and convergence status in an output file (\texttt{out.json}).

The initial sampling phase employs a quasi-random Sobol sequence with \(N_{\text{init}} = 8\) points to explore the design space efficiently and populate the training set for the Gaussian process (GP) surrogates. Two independent GP models are constructed: one for \(T_m\) (the objective function) and another for \(NO_x\) (the probabilistic constraint). Each GP uses a Matérn 2.5 kernel with automatic relevance determination (ARD) and Gamma priors on the lengthscales and output variance. The likelihood variance is bounded below by a minimum noise level of \(10^{-4}\) to prevent numerical instabilities during fitting. Input variables are normalized to \([0,1]^3\), and outputs are standardized for numerical conditioning. Model hyperparameters are optimized via maximization of the exact marginal log-likelihood, and warm-starting is applied at each iteration by reusing the previous GP state to accelerate convergence.

At each iteration of the optimization, the constrained noisy expected improvement \(\alpha_{\text{qNEI}}^{(\text{constr})}\) is evaluated across the design space using a Sobol–QMC normal sampler with 256 Monte Carlo samples. The candidate point \(\mathbf{x}^*\) is obtained by maximizing the cost-aware penalized acquisition function defined in Equation~\eqref{eq:acquisition_penalized}. Numerical optimization of the acquisition function is carried out using the multi-start routine \texttt{optimize\_acqf} from BoTorch, with 12 restarts and 512 raw samples to ensure global exploration.

Each new candidate is automatically submitted to ANSYS Workbench, and the returned results are logged in \texttt{bo\_history.csv}. This file records, for every iteration, the design variables (\(h,\,l,\,s\)), the derived fidelity indicator \(z\), the mean temperature \(T_m\), NO$_x$ emissions, mesh size, wall-clock time, acquisition value, and GP hyperparameters (lengthscales, noise, and output scale). This continuous record allows real-time monitoring of the optimization progress and post-analysis of the model behavior.

Overall, the workflow integrates CFD simulations, surrogate modeling, and adaptive Bayesian learning into a fully automated closed loop. By coupling BoTorch and GPyTorch with ANSYS Workbench through Python automation, the framework ensures reproducible and scalable multi-fidelity optimization under realistic computational constraints.

\clearpage 

\section{Results and Discussion}

This section presents (i) the convergence and consistency of the Gaussian Process (GP) surrogate models for mean temperature (\(T_m\)) and nitrogen oxides (\(NO_x\)), (ii) the spatial predictions and associated uncertainties, and (iii) the relationship between reactor geometry \((h,l)\), fidelity \(Z\), performance, and computational cost. All figures correspond to the results of the multi-fidelity Bayesian optimization procedure described in Methodology.

\subsection{GP convergence and spatial prediction}

The evolution of the GP hyperparameters encodes how the surrogate balances smoothness, signal variance, and observation noise as new data are acquired~\citep{Rasmussen2006,Garnett2023}. 
In GP models, smaller lengthscales imply sharper variation and stronger sensitivity to the associated input dimension, whereas larger lengthscales indicate smoother trends and lower sensitivity. 
To make both the numerical convergence and the physical interpretability explicit,side-by-side, the raw lengthscales and their normalized ARD importance, for \(T_m\) and \(NO_x\).

\begin{figure}[!h]
  \centering
  \begin{subfigure}{0.48\linewidth}
    \centering
    \includegraphics[width=\linewidth]{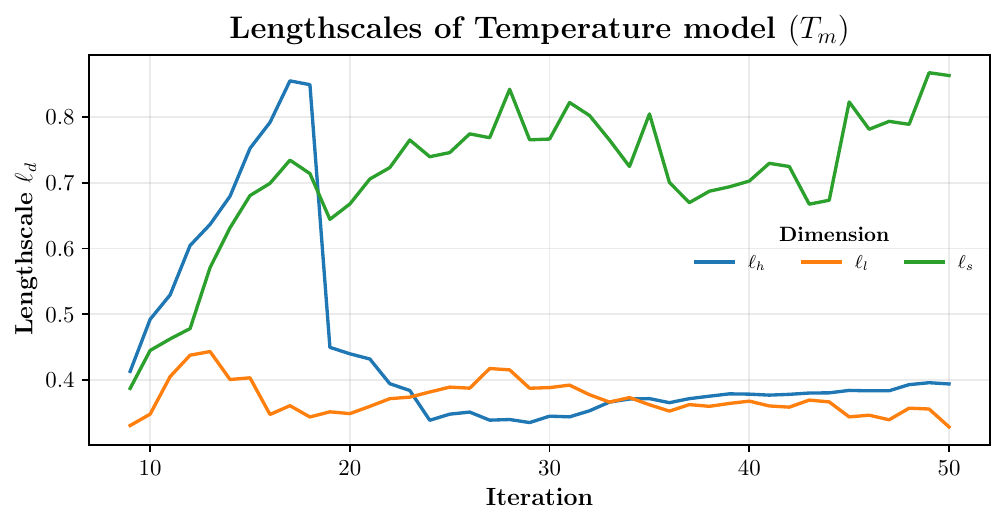}
    \caption{Lengthscales \(\ell_h,\ell_l,\ell_s\) para \(T_m\).}
    \label{fig:len_T}
  \end{subfigure}\hfill
  \begin{subfigure}{0.48\linewidth}
    \centering
    \includegraphics[width=\linewidth]{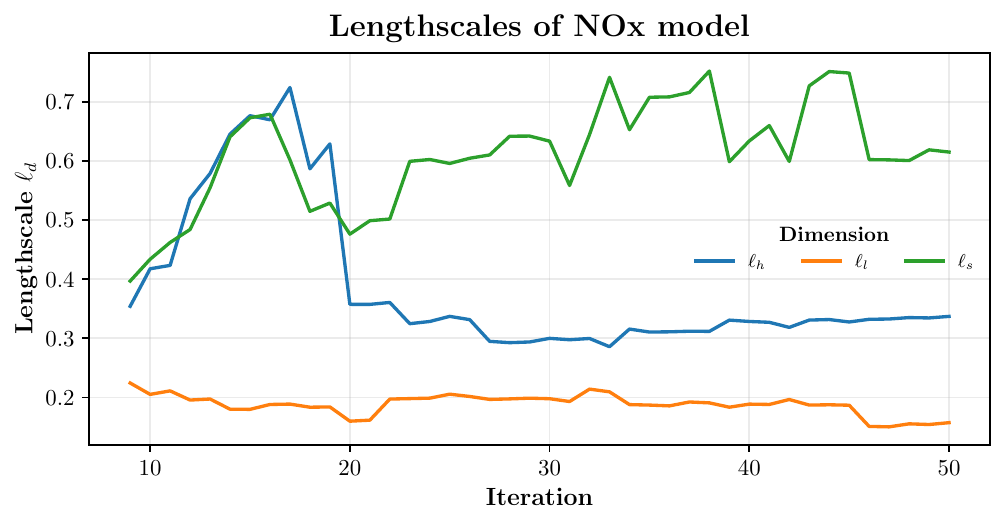}
    \caption{Lengthscales \(\ell_h,\ell_l,\ell_s\) para \(NO_x\).}
    \label{fig:len_NOx}
  \end{subfigure}
  \caption{Evolution of the GP lengthscales for \(T_m\) (left) and \(NO_x\) (right). The stabilization after approximately 30 iterations indicates numerical convergence of the hyperparameters.}
  \label{fig:lengthscales_pair}
\end{figure}

\begin{figure}[H]
  \centering
  \begin{subfigure}{0.48\linewidth}
    \centering
    \includegraphics[width=\linewidth]{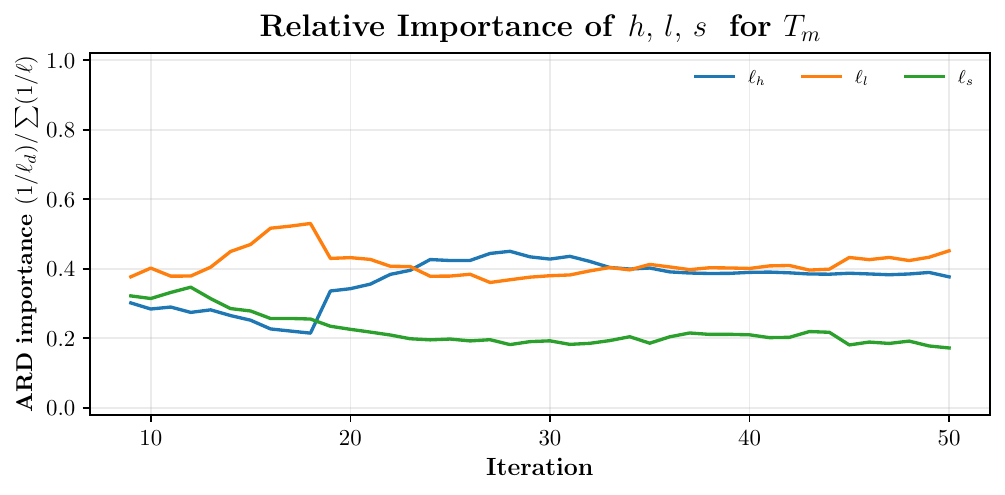}
    \caption{Normalized ARD importance for \(T_m\).}
    \label{fig:ard_T}
  \end{subfigure}\hfill
  \begin{subfigure}{0.48\linewidth}
    \centering
    \includegraphics[width=\linewidth]{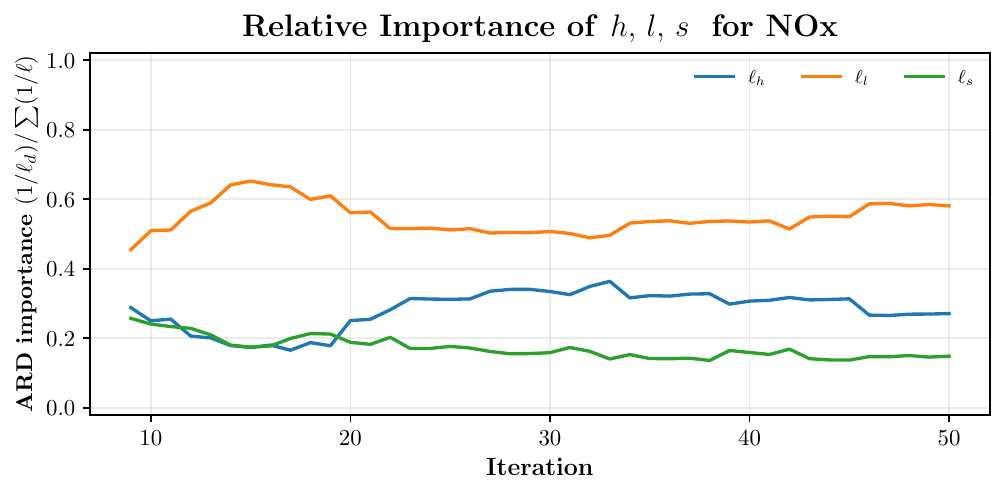}
    \caption{Normalized ARD importance for \(NO_x\).}
    \label{fig:ard_NOx}
  \end{subfigure}
  \caption{Normalized inverse lengthscales \((1/\ell_d)/\sum(1/\ell)\) for \(T_m\) (left) and \(NO_x\) (right), highlighting relative sensitivity to \(h,l,s\).}
  \label{fig:ard_pair}
\end{figure}

From \Cref{fig:lengthscales_pair}, the characteristic lengthscales of both surrogates stabilize after approximately 30 iterations, indicating a consistent posterior landscape and numerically stable training. 
The ARD panels, \Cref{fig:ard_pair}, translate those trends into relative importance: for temperature, \(l\) and \(s\) dominate (consistent with the influence of flow development and mesh resolution on heat transfer), whereas for \(NO_x\) the height \(h\) becomes comparatively more influential (aligned with near-wall mixing and residence-time effects). 
Presenting both views together conveys (i) that the GP hyperparameters converge, and (ii) \textit{why} the model behaves as observed along the design dimensions~\citep{Bishop2006}.

\begin{figure}[H]
  \centering
  \begin{subfigure}{0.48\textwidth}
    \centering
    \includegraphics[width=\linewidth]{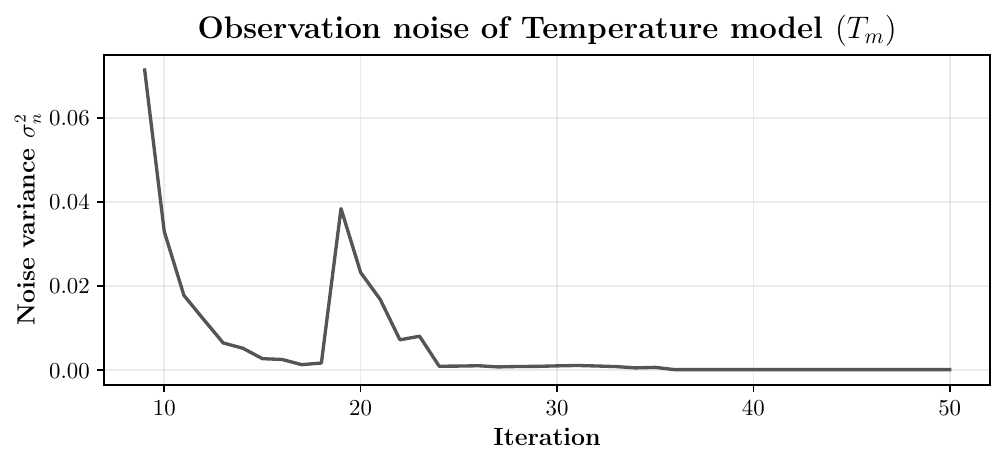}
    \caption{Observation-noise variance of the \(T_m\) model.}
    \label{fig:noise_mll_a}
  \end{subfigure}\hfill
  \begin{subfigure}{0.48\textwidth}
    \centering
    \includegraphics[width=\linewidth]{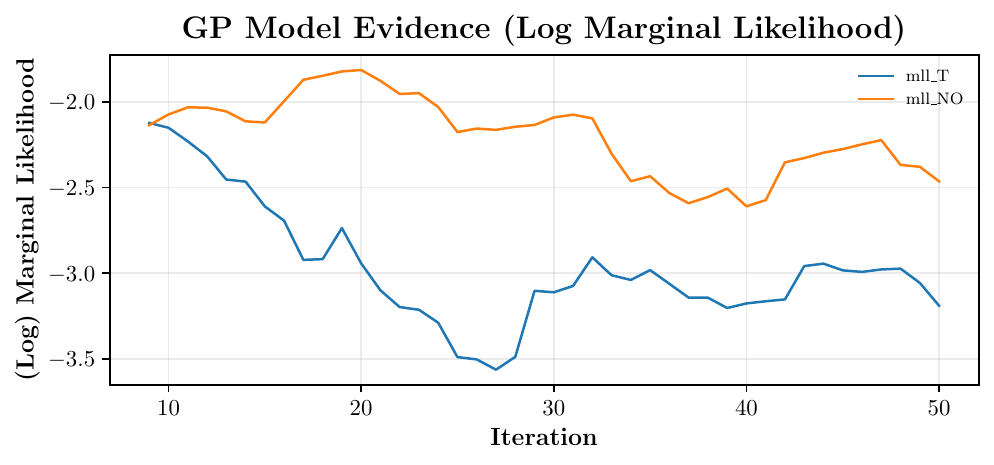}
    \caption{Log marginal likelihood for \(T_m\) and \(NO_x\).}
    \label{fig:noise_mll_b}
  \end{subfigure}
  \caption{Noise variance and model evidence confirm numerically stable training.}
  \label{fig:noise_mll}
\end{figure}

In parallel, the observation-noise, Figure~\ref{fig:noise_mll_a}, variance and the model evidence evolve coherently.
The noise variance \(\sigma_n^2\) of the temperature model decreases to values below \(0.05~\mathrm{K}^2\), reflecting high simulation fidelity, while the (log) marginal likelihood, Figure~\ref{fig:noise_mll_b}, increases throughout training, evidencing an improved trade-off between data fit and complexity~\citep{MacKay1992,Rasmussen2006}. 
A transient increase in the noise variance was observed around iteration~20, 
Figure~\ref{fig:noise_mll_a}, coinciding with the addition of a new data point 
that introduced higher local variability in the temperature response. 
This temporary rise indicates that the optimizer momentarily inflated the 
observation-noise term to reconcile the new information with the existing posterior. 
As more samples were added, the noise variance quickly decayed and the marginal 
likelihood stabilized, confirming the robustness of the learned surrogate.

The concurrent stabilization of these quantities supports the robustness of the learned surrogates.

\begin{figure}[H]
  \centering
  \begin{subfigure}{0.48\textwidth}
    \centering
    \includegraphics[width=\linewidth]{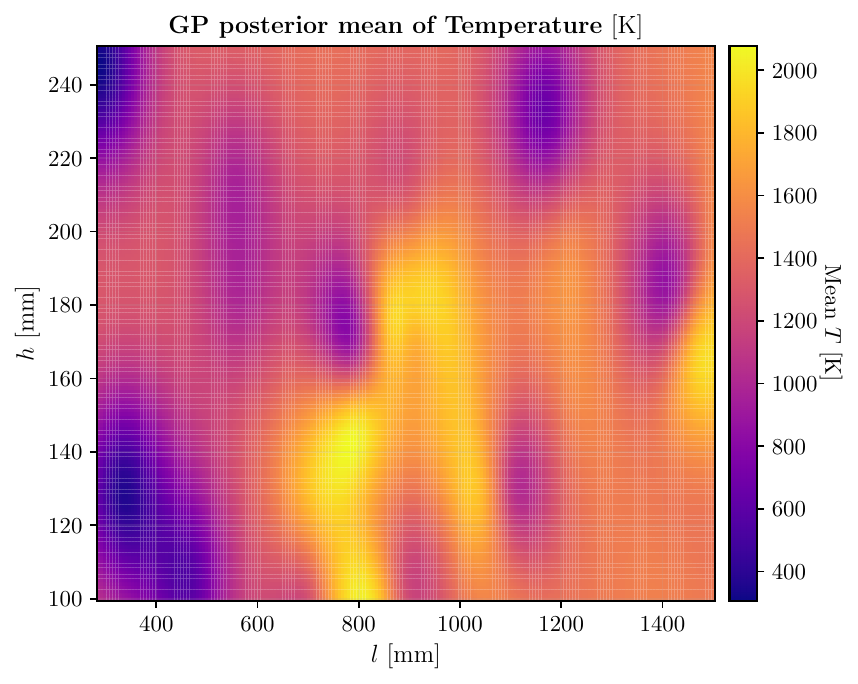}
    \caption{Posterior mean of \(T_m\) over \((l,h)\).}
    \label{fig:gp_mean_T}
  \end{subfigure}\hfill
  \begin{subfigure}{0.48\textwidth}
    \centering
    \includegraphics[width=\linewidth]{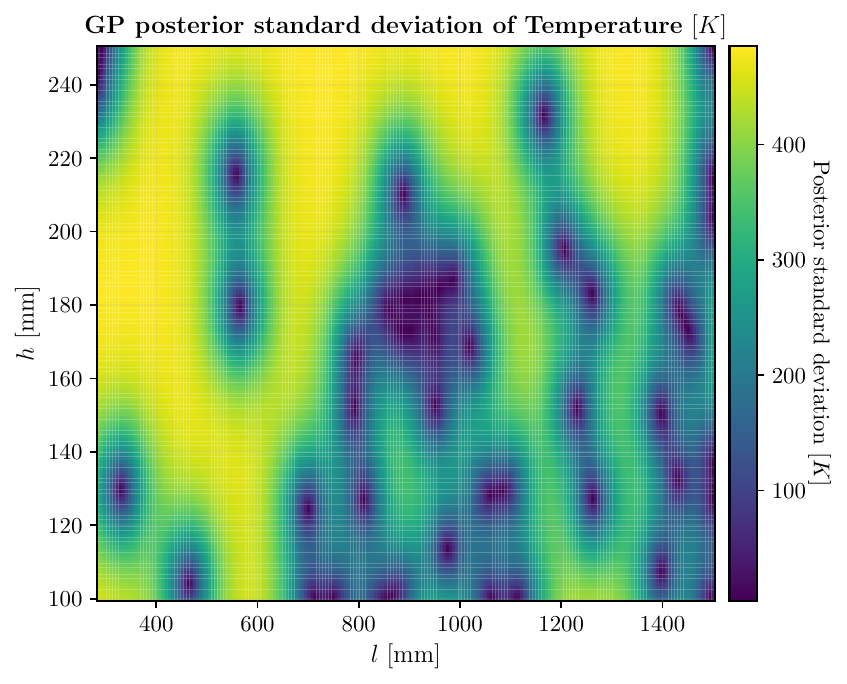}
    \caption{Posterior standard deviation of \(T_m\).}
    \label{fig:gp_std_T}
  \end{subfigure}
  \caption{Spatial prediction and uncertainty of the temperature surrogate.}
\end{figure}

The GP posterior mean and standard deviation maps for \(T_m\) are shown in \Cref{fig:gp_mean_T,fig:gp_std_T}. 
The mean map reproduces expected gradients across the \((l,h)\) plane, while the standard deviation highlights higher uncertainty near domain boundaries and in sparsely sampled low-fidelity regions (\(Z<0.4\)), a pattern consistent with Bayesian learning~\citep{Snoek2012}. 
Accordingly, it can be observed that during the optimization process, the region defined by reactor length \(l \in [800,\,1000]\)~mm and height \(h \in [100,\,180]\)~mm was recognized as a promising design zone for maximizing the mean temperature, thereby explaining the lower posterior standard deviation in this part of the domain.
These maps provide actionable guidance for where high-fidelity evaluations would most effectively reduce  uncertainty.

\subsection{Strategy of Penalized Acquisition,Cost Modeling and Time Trade-Off}

\begin{figure}[h]
    \centering
    \includegraphics[width=0.8\linewidth]{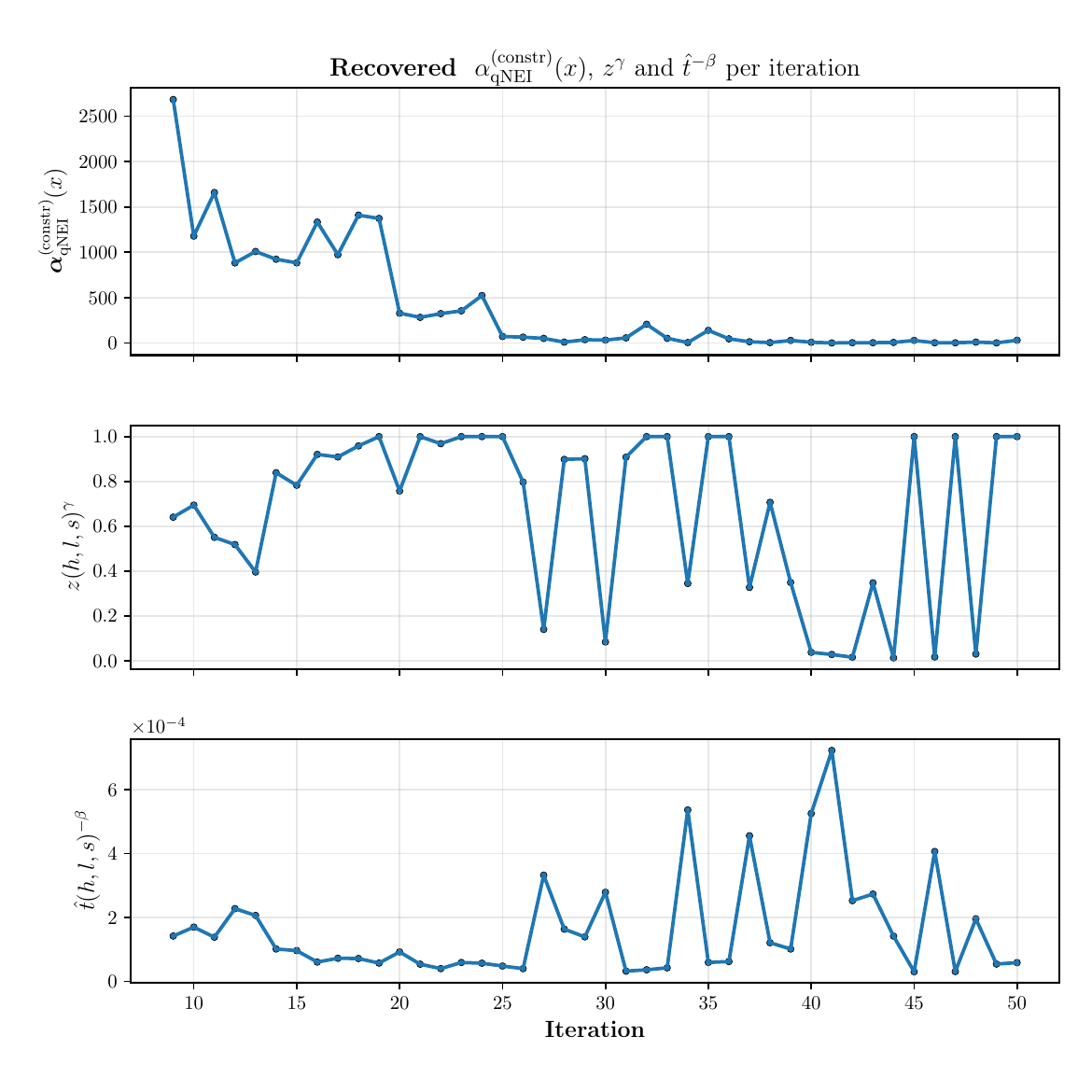}
    \caption{Evolution of the constrained qNEI acquisition term and penalization factors along the optimization iterations. 
    The dominance of low-fidelity evaluations in early iterations reflects exploration, while high-fidelity evaluations become prevalent during exploitation.}
    \label{fig:penalizadores}
\end{figure}

The proposed optimization framework couples the constrained qNEI acquisition function with two penalization terms that account for both model fidelity and computational time. Figure~\ref{fig:penalizadores} illustrates the evolution of these components throughout the optimization process: the baseline acquisition term $\alpha_{\mathrm{qNEI}}^{(\mathrm{constr})}(x)$, the fidelity penalization $z(h,l,s)^{\gamma}$, and the time penalization $\hat{t}(h,l,s)^{-\beta}$.

At the beginning of the optimization, the \textit{qNEI} term dominates, leading the optimizer to explore broad regions of the design space through low-fidelity simulations ($Z<0.4$). As the process evolves, the combined penalization terms progressively shift the search behavior toward exploitation, focusing computational effort on the most promising high-fidelity regions. This adaptive balance allows the optimizer to “spend” computational resources strategically: early iterations rely on cheaper low-fidelity evaluations to identify global trends, while later iterations invest in expensive high-fidelity runs once convergence is near.

This dynamic exploration–exploitation transition is a key feature of the multi-fidelity Bayesian approach, enabling convergence with far fewer expensive CFD simulations compared to traditional single-fidelity optimization strategies.

\begin{figure}[H]
    \centering
    \includegraphics[width=0.8\linewidth]{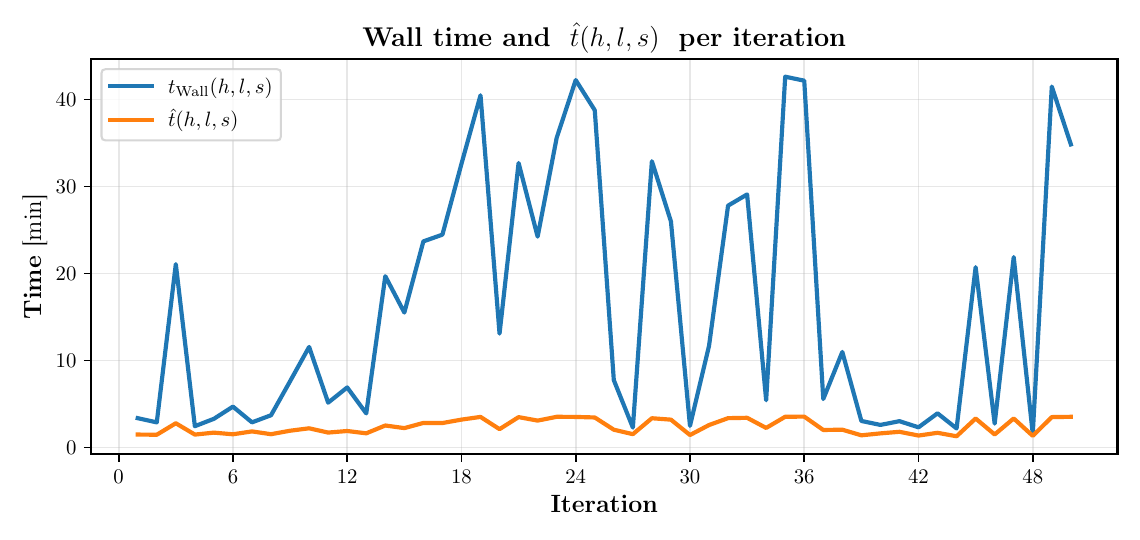}
    \caption{Comparison between the modeled time $\hat{t}(h,l,s)$ and the real wall time $t_{\mathrm{Wall}}(h,l,s)$ per iteration. 
    The modeled curve captures the overall trend, providing sufficient directional feedback for the penalized acquisition function.}
    \label{fig:walltime_iter}
\end{figure}

The relationship between modeled and real wall times provides further insight into how the optimizer manages computational resources. 
Figure~\ref{fig:walltime_iter} compares the predicted simulation time $\hat{t}(h,l,s)$ with the actual measured wall time $t_{\mathrm{Wall}}(h,l,s)$ per iteration.
Although discrepancies are visible,particularly during early calibration, the model preserves the correct monotonic trend: when the real wall time increases, the predicted time also increases.

\begin{figure}[H]
    \centering
    \includegraphics[width=0.8\linewidth]{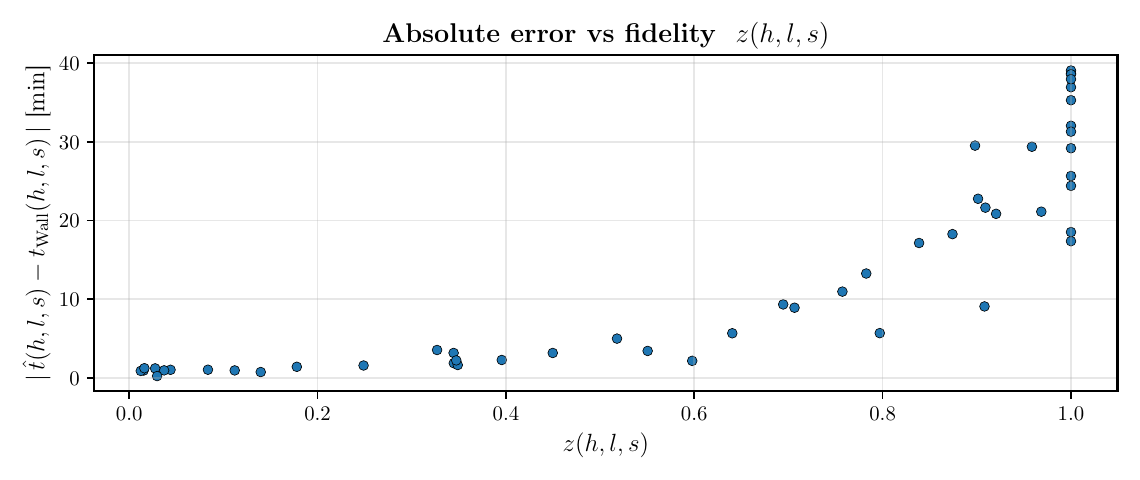}
    \caption{Absolute difference between modeled and real wall time as a function of fidelity $z(h,l,s)$. 
    Despite increasing uncertainty for high fidelities, the model maintains the correct qualitative trend.}
    \label{fig:absdiff_z}
\end{figure}

Importantly, the objective of this surrogate model is not to achieve perfect regression accuracy, but to provide reliable directional feedback to the acquisition function. 
As long as $\hat{t}$ correctly distinguishes between cheap and expensive simulations, the optimization algorithm can adjust its sampling strategy accordingly.
This concept is reinforced in Fig.~\ref{fig:absdiff_z}, which shows that the absolute error between $\hat{t}$ and $t_{\mathrm{Wall}}$ grows with fidelity, yet maintains consistent behavior.

\begin{figure}[H]
    \centering
    \includegraphics[width=0.8\linewidth]{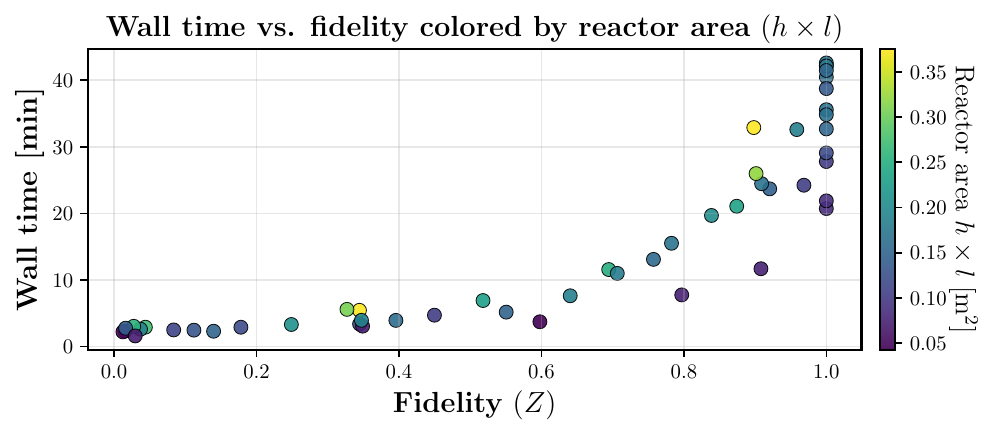}
    \caption{Wall time as a function of fidelity $Z$, colored by reactor area $(h\times l)$. 
    Low-fidelity simulations ($Z<0.4$) are significantly faster, highlighting the computational efficiency of the multi-fidelity approach.}
    \label{fig:wallxz}
\end{figure}

Finally, the wall time–fidelity map (Fig.~\ref{fig:wallxz}) quantifies the computational
advantage achieved by the proposed approach. Low-fidelity evaluations
($Z<0.4$) are roughly an order of magnitude faster than full-fidelity ones,
enabling the optimizer to learn the search landscape efficiently before committing
to expensive CFD runs. In contrast, a traditional CFD-based optimization
relying solely on high-fidelity simulations would require prohibitive computational
time for comparable convergence.

To better illustrate this advantage, Table~\ref{tab:time_comparison} provides an
estimated comparison between the total computational cost of the proposed
multi-fidelity optimization and a hypothetical single-fidelity approach using
only $Z=1$. The estimation is based on the wall-time data from
Fig.~\ref{fig:walltime_iter}, assuming $N=50$ iterations and average wall times
of 35~min for full fidelity and 15~min for the mixed-fidelity runs.

\begin{table}[h]
    \centering
    \caption{Estimated computational cost of the optimization under different fidelity strategies.}
    \label{tab:time_comparison}
    \vspace{4pt}
    \resizebox{0.9\linewidth}{!}{%
    \begin{tabular}{lccc}
        \hline
        \textbf{Strategy} & \textbf{Average wall time [min]} & \textbf{Total time [min]} & \textbf{Total time [h]} \\
        \hline
        Single-fidelity ($Z=1$) & 35 & 1750 & 29.2 \\
        Multi-fidelity (proposed) & 15 & 750 & 12.5 \\
        \hline
        \textbf{Time reduction} & --- & \textbf{$-1000$} & \textbf{$\approx 57\%$} \\
        \hline
    \end{tabular}%
    }
\end{table}

In practical terms, performing all iterations using full-fidelity CFD would
require approximately $T_{\mathrm{only\,HF}} \approx 1750~\text{min}$ (about 29~hours).
In contrast, the proposed multi-fidelity strategy completed the same number
of iterations in $T_{\mathrm{MF}} \approx 750~\text{min}$ (about 12.5~hours), corresponding
to a total computational saving of approximately 57\%.

\subsection{Optimization trajectory and physico–computational trade-offs}
\label{sec:traj_tradeoffs}

\begin{figure}[H]
  \centering
  \includegraphics[width=\textwidth]{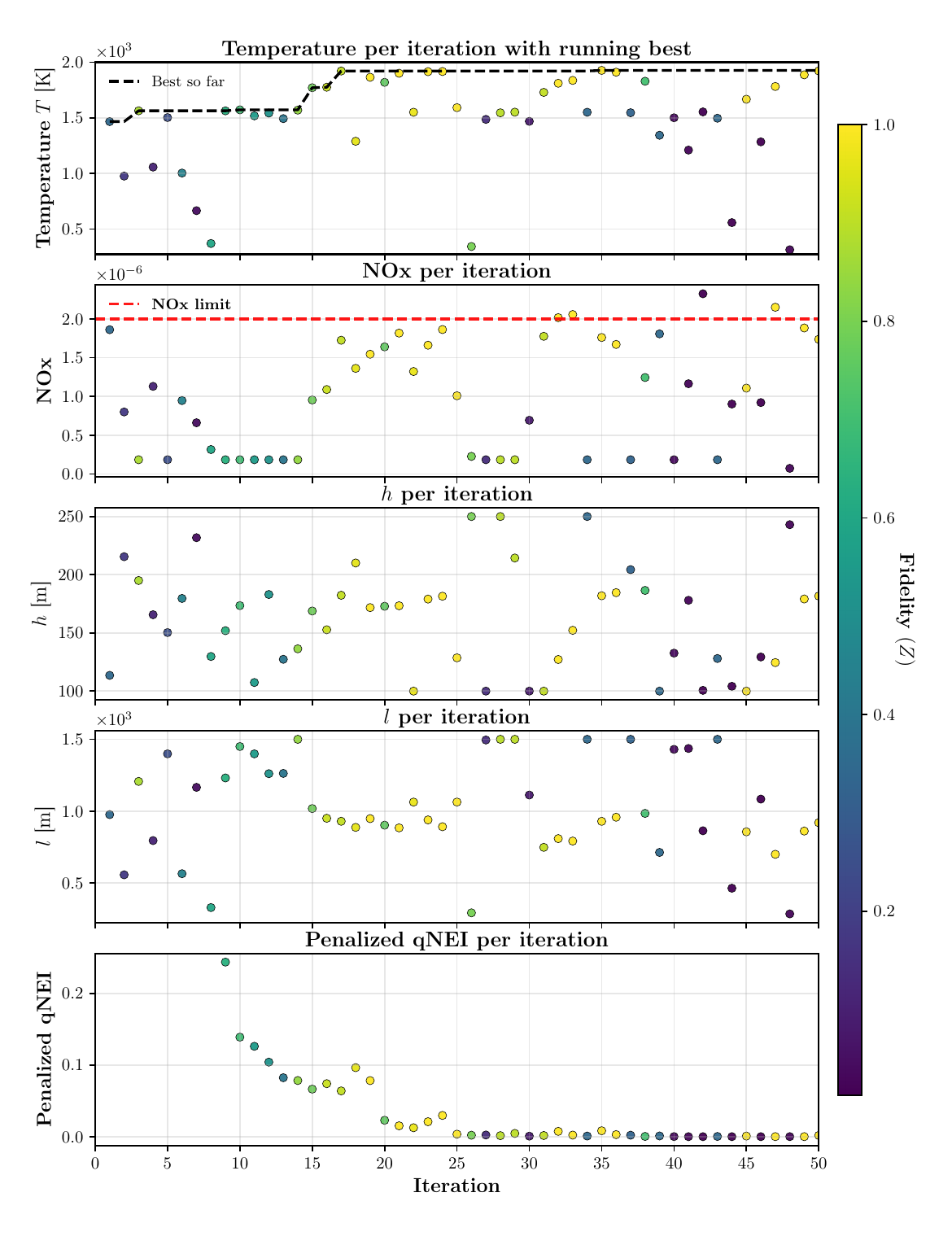}
  \caption{Best-so-far temperature, geometry $(h,l)$ path, and fidelity $Z$ along the optimization.}
  \label{fig:best_so_far}
\end{figure}

Trajectory of improvement and endogenous fidelity scheduling Figure~\ref{fig:best_so_far} condenses the search dynamics: the running best
temperature increases monotonically as the acquisition drives evaluations toward
progressively more promising geometries, while the fidelity $Z$ rises toward the
end of the run. This endogenic increase of $Z$ is expected under a cost-aware
policy: early iterations use low-cost, low-$Z$ probes to map global trends and
shrink posterior uncertainty; as information concentrates around a small set of
candidates, paying for higher fidelity becomes rational because the expected gain
per wall-time grows. In practice, increasing $Z$ (i.e., refining $s$) sharpens
the resolution of near-wall gradients, recirculation pockets, and flame–shear
interactions that underpin hot spots, so the GP posterior for $T_m$ tends to lift
when evaluated at finer meshes near a good design. Crucially, this migration is
constrained by the NO$_x$ feasibility probability, preventing high-temperature,
high-$Z$ moves that would breach the emission limit.

\begin{figure}[H]
  \centering
  \includegraphics[width=0.62\textwidth]{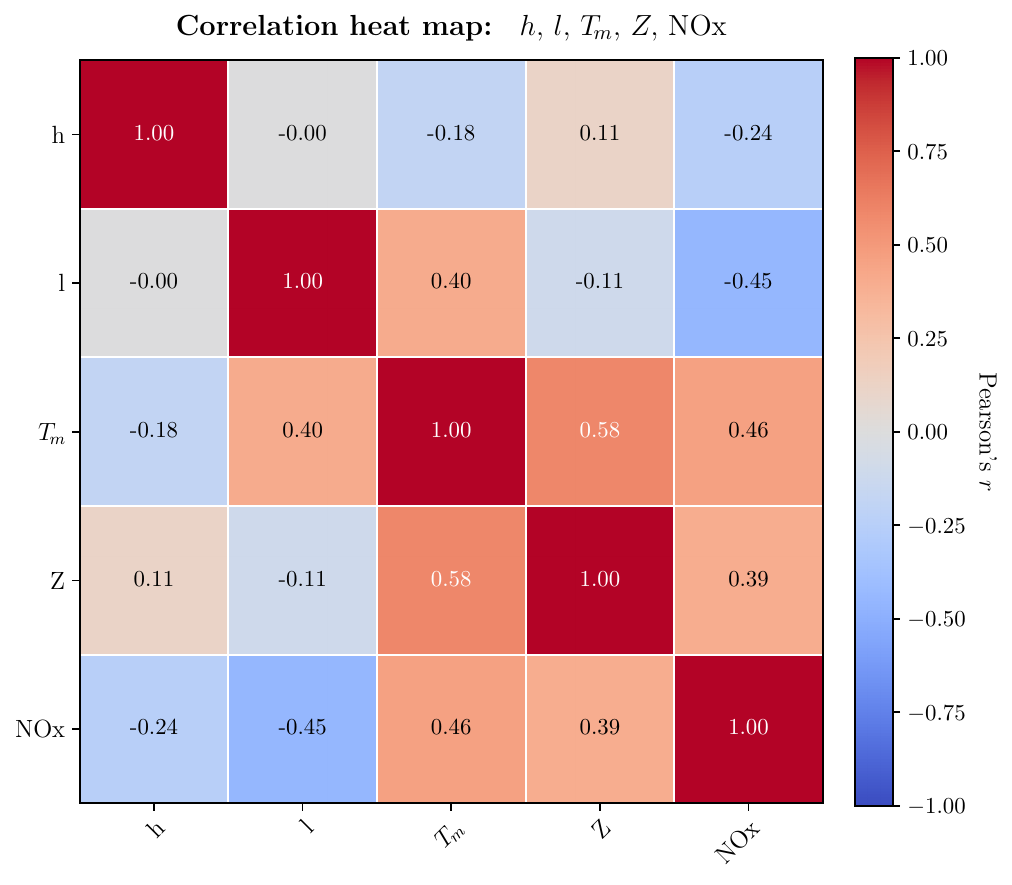}
  \caption{Correlation map among $h$, $l$, $T_m$, $Z$, and NO$_x$. Positive $T_m$–$l$ and $T_m$–$Z$,
           and negative NO$_x$–$l$, are consistent with length-enabled mixing and fidelity-enabled
           resolution of hot structures, under an NO$_x$ constraint.}
  \label{fig:corr_heatmap}
\end{figure}

Correlation structure: what the optimizer ``learned'' about geometry, fidelity, and responses can be seen the Pearson heat map in Figure~\ref{fig:corr_heatmap} summarizes the global couplings
observed during the run across height $h$, length $l$, mean temperature $T_m$,
fidelity $Z$, and NO$_x$.

$T_m$ shows a positive association with reactor length $l$ and a
\emph{slightly negative} association with height $h$. A longer combustor allows
flow development and mixing to mature downstream of the injection zone; coherent
recirculation and shear-layer entrainment stabilize heat-release structures that
raise the domain-average $T_m$. Increasing $h$, by contrast, redistributes
thermal gradients normal to the wall and softens confinement, which can dilute
peak thermal regions and slightly depress $T_m$ at fixed inlet conditions.
$T_m$ correlates positively with $Z$: finer meshes capture sharper
scalar and velocity gradients (reduced numerical diffusion), revealing stronger
localized heat-release that lifts the posterior mean of $T_m$ near promising
designs. This is beneficial for the optimizer: the acquisition naturally
tends to increase $Z$ where both expected improvement and feasibility remain high.
NO$_x$ exhibits a positive correlation with $T_m$ (thermal pathway),
but tends to decrease with $l$, suggesting that extended mixing and
distributed heat-release temper the peak temperatures that drive Zeldovich NO.
Together, these signs explain why the search gravitates toward larger $l$,
moderate $h$, and higher $Z$ only in regions where the NO$_x$ constraint
is still likely to hold.

\begin{figure}[H]
  \centering
  \includegraphics[width=0.75\textwidth]{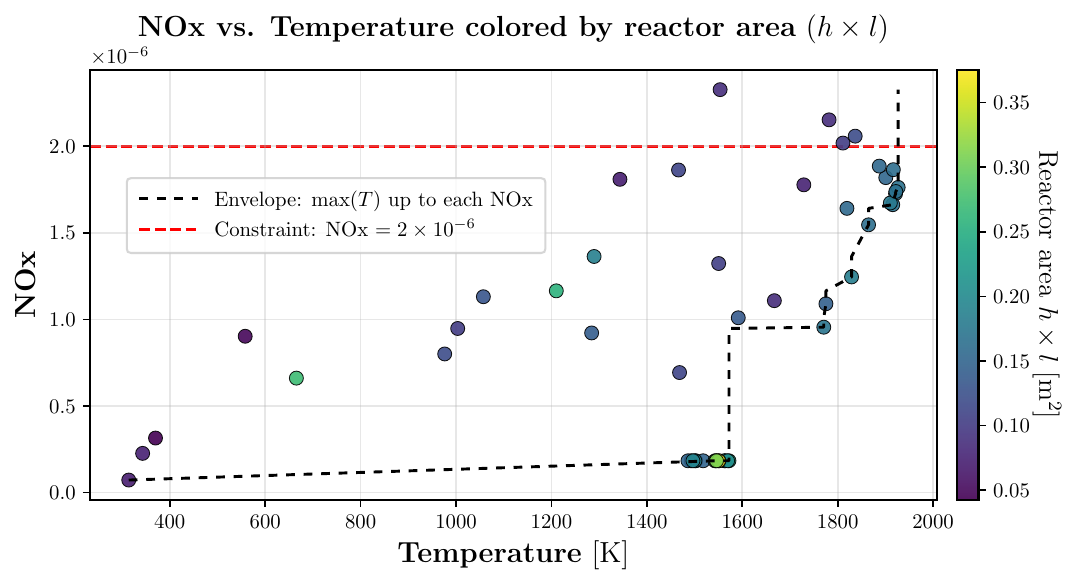}
  \caption{NO$_x$ versus $T_m$, colored by reactor area $h \times l$. The cumulative maximum-temperature envelope highlights two regimes: an initial region where $T_m$ increases with only mild NO$_x$ growth, followed by a steep rise in emissions beyond the \emph{elbow}. Favorable trade-offs cluster at larger areas, mainly longer $l$, which enhance flow development and promote higher mean temperatures without exceeding the NO$_x = 2 \times 10^{-6}$ cap. From an environmental standpoint, operation near the elbow offers a balanced compromise between thermal efficiency and emission control.}

  \label{fig:nox_vs_T}
\end{figure}

Figure~\ref{fig:nox_vs_T} plots NO$_x$ against $T_m$, colored by reactor area $h\times l$. Two regimes emerge. 
The fist is mixing–boosted temperature with mild NO$_x$ growth, for intermediate $T_m$, designs with larger $h\times l$ (driven mainly by $l$) cluster along a favorable band where $T_m$ increases substantially while NO$_x$ rises slowly. This aligns with the correlation map, where $l$ correlates positively with $T_m$ and negatively with NO$_x$, indicating that extended mixing and flow development raise the mean temperature without proportionally increasing peak temperatures. In other words, $l$ improves thermal utilization more than it intensifies the local hot spots that trigger thermal NO$_x$.

The second zone is the thermal NO$_x$ dominated escalation, beyond the envelope “elbow,” small gains in $T_m$ lead to sharp increases in NO$_x$, consistent with the strong temperature sensitivity of the extended Zeldovich mechanism once high-$T$ pockets and residence time become significant. In this region, any further push in $T_m$ tends to concentrate heat release and elongate high-$T$ zones, steepening the NO$_x$ slope.

This picture is consistent with the surrogate diagnostics. The GP mean of $T_m(h,l)$ shows a ridge that grows with $l$, while the posterior uncertainty concentrates near that ridge; the optimizer therefore exploits larger $l$ where expected improvement and feasibility are both high, then increases fidelity $Z$ to lock in gains once the feasible band is identified. ARD analyses further show $l$ as the most influential variable for $T_m$, with $h$/$s$ playing a secondary role, whereas NO$_x$ becomes more sensitive once the design approaches the high $T$ ridge precisely where the slope in Figure~\ref{fig:nox_vs_T} steepens. 

Design takeaway. From an environmental standpoint, the most desirable operating zone lies near the elbow of the $T_m$–NO$_x$ curve, where satisfactory mean temperatures can be achieved without a significant rise in NO$_x$. In this region, designs with moderately large $l$ and reactor area offer an efficient balance between combustion performance and emission control. Beyond this point, further temperature gains lead to a disproportionate increase in NO$_x$, suggesting that stricter attention to the emission constraint becomes necessary.

\clearpage 
\section{Conclusions}

This work introduced a multi-fidelity Bayesian optimization for the geometric design of a non-premixed methane–hydrogen burner, coupling high-fidelity CFD simulations with Gaussian-process surrogates and a cost-aware acquisition strategy. The approach demonstrated that efficient design exploration of reacting-flow systems can be achieved without relying exclusively on costly full-resolution simulations.

The proposed formulation integrates three key components: (i) a continuous fidelity index $Z(h,l,s)$ based on mesh density, allowing smooth transitions between coarse and fine CFD evaluations; (ii) a calibrated runtime model $\hat{t}(h,l,s)$ that captures the nonlinear cost scaling with geometry and resolution; and (iii) a constrained, cost-penalized acquisition function $\alpha_{\mathrm{qNEI}}^{(\mathrm{constr})}$ that simultaneously drives performance improvement, enforces NO$_x$ feasibility, and limits computational expenditure. Together, these elements enable the optimizer to allocate resources adaptively, favoring inexpensive low-$Z$ simulations for global exploration and gradually increasing fidelity as convergence is approached.

The results confirmed the numerical stability and physical interpretability of the Gaussian-process surrogates. Hyperparameters converged  after about 30 iterations, and the ARD analysis revealed that reactor length $l$ was the dominant design variable influencing both temperature and emissions. The correlation and sensitivity maps established that longer combustors enhance flow development and mixing, promoting higher mean temperatures while mitigating peak local temperatures responsible for thermal NO$_x$.

The optimizer naturally identified two operating regimes. In the first, moderate increases in $T_m$ occur with only mild NO$_x$ growth, reflecting mixing-enhanced performance at larger $l$. Beyond the “elbow” of the $T_m$–NO$_x$ curve, however, further temperature gains cause a sharp rise in emissions, consistent with the onset of Zeldovich-driven thermal NO$_x$. From an environmental standpoint, the most favorable operating zone lies near this transition region, where high thermal efficiency can be achieved without exceeding the $2\times10^{-6}$ NO$_x$ cap.

Computationally, the multi-fidelity Bayesian optimization  convergence with approximately $57\%$ lower total wall time compared to a hypothetical single-fidelity campaign, confirming the benefits of integrating fidelity-aware penalization and runtime modeling into the acquisition function. The resulting workflow, implemented through a Python/BoTorch–ANSYS interface, proved robust and fully automated, offering a practical route to cost-effective optimization of high-dimensional CFD-driven designs.

\clearpage 
\bibliography{refs}

\end{document}